\documentclass[twoside,twocolumn,9pt]{article}
\usepackage{extsizes}
\usepackage[super,sort&compress,comma]{natbib} 
\usepackage[version=3]{mhchem}
\usepackage[left=1.5cm, right=1.5cm, top=1.785cm, bottom=2.0cm]{geometry}
\usepackage{balance}
\usepackage{mathptmx}
\usepackage{sectsty}
\usepackage{graphicx} 
\usepackage{lastpage}
\usepackage[format=plain,justification=justified,singlelinecheck=false,font={stretch=1.125,small,sf},labelfont=bf,labelsep=space]{caption}
\usepackage{float}
\usepackage{fancyhdr}
\usepackage{fnpos}
\usepackage[english]{babel}
\addto{\captionsenglish}{%
  
}
\usepackage{array}
\usepackage{droidsans}
\usepackage{charter}
\usepackage[T1]{fontenc}
\usepackage[usenames,dvipsnames]{xcolor}
\usepackage{setspace}
\usepackage[compact]{titlesec}
\usepackage{hyperref}
\usepackage{xspace}
\usepackage{siunitx}
\usepackage{booktabs}
\usepackage{multirow}
\usepackage{comment}
\def\bnNaph{$\mathbf{BN}_{\mathbf{Naph}}$\xspace}
\def\Naph{naphthalene\xspace}
\def\Benz{benzene\xspace}

\def\bnBenz{$\mathbf{BN}_{\mathbf{Benz}}$\xspace}

\usepackage{epstopdf}%This line makes .eps figures into .pdf - please comment out if not required.

\definecolor{cream}{RGB}{222,217,201}

\begin{document}

\pagestyle{fancy}
\thispagestyle{plain}
\fancypagestyle{plain}{
%%%HEADER%%%
\renewcommand{\headrulewidth}{0pt}
}
%%%END OF HEADER%%%

%%%PAGE SETUP - Please do not change any commands within this section%%%
\makeFNbottom
\makeatletter
\renewcommand\LARGE{\@setfontsize\LARGE{15pt}{17}}
\renewcommand\Large{\@setfontsize\Large{12pt}{14}}
\renewcommand\large{\@setfontsize\large{10pt}{12}}
\renewcommand\footnotesize{\@setfontsize\footnotesize{7pt}{10}}
\makeatother

\renewcommand{\thefootnote}{\fnsymbol{footnote}}
\renewcommand\footnoterule{\vspace*{1pt}% 
\color{cream}\hrule width 3.5in height 0.4pt \color{black}\vspace*{5pt}} 
\setcounter{secnumdepth}{5}

\makeatletter 
\renewcommand\@biblabel[1]{#1}            
\renewcommand\@makefntext[1]% 
{\noindent\makebox[0pt][r]{\@thefnmark\,}#1}
\makeatother 
\renewcommand{\figurename}{\small{Fig.}~}
\sectionfont{\sffamily\Large}
\subsectionfont{\normalsize}
\subsubsectionfont{\bf}
\setstretch{1.125} %In particular, please do not alter this line.
\setlength{\skip\footins}{0.8cm}
\setlength{\footnotesep}{0.25cm}
\setlength{\jot}{10pt}
\titlespacing*{\section}{0pt}{4pt}{4pt}
\titlespacing*{\subsection}{0pt}{15pt}{1pt}
%%%END OF PAGE SETUP%%%

%%%FOOTER%%%
\fancyfoot{}
\fancyfoot[LO,RE]{\vspace{-7.1pt}\includegraphics[height=9pt]{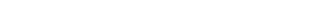}}
\fancyfoot[CO]{\vspace{-7.1pt}\hspace{11.9cm}\includegraphics{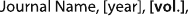}}
\fancyfoot[CE]{\vspace{-7.2pt}\hspace{-13.2cm}\includegraphics{head_foot/RF}}
\fancyfoot[RO]{\footnotesize{\sffamily{1--\pageref{LastPage} ~\textbar  \hspace{2pt}\thepage}}}
\fancyfoot[LE]{\footnotesize{\sffamily{\thepage~\textbar\hspace{4.65cm} 1--\pageref{LastPage}}}}
\fancyhead{}
\renewcommand{\headrulewidth}{0pt} 
\renewcommand{\footrulewidth}{0pt}
\setlength{\arrayrulewidth}{1pt}
\setlength{\columnsep}{6.5mm}
\setlength\bibsep{1pt}
%%%END OF FOOTER%%%

%%%FIGURE SETUP - please do not change any commands within this section%%%
\makeatletter 
\newlength{\figrulesep} 
\setlength{\figrulesep}{0.5\textfloatsep} 

\newcommand{\topfigrule}{\vspace*{-1pt}% 
\noindent{\color{cream}\rule[-\figrulesep]{\columnwidth}{1.5pt}} }

\newcommand{\botfigrule}{\vspace*{-2pt}% 
\noindent{\color{cream}\rule[\figrulesep]{\columnwidth}{1.5pt}} }

\newcommand{\dblfigrule}{\vspace*{-1pt}% 
\noindent{\color{cream}\rule[-\figrulesep]{\textwidth}{1.5pt}} }

\makeatother
%%%END OF FIGURE SETUP%%%

%%%TITLE, AUTHORS AND ABSTRACT%%%
\twocolumn[
  \begin{@twocolumnfalse}
%{\includegraphics[height=30pt]{head_foot/PCCP}\hfill\raisebox{0pt}[0pt][0pt]{\includegraphics[height=55pt]{head_foot/RSC_LOGO_CMYK}}\\[1ex]
%\includegraphics[width=18.5cm]{head_foot/header_bar}}\par
\vspace{1em}
\sffamily
\begin{tabular}{m{4.5cm} p{13.5cm} }

\includegraphics{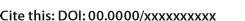} & \noindent\LARGE{\textbf{Asymmetric Planar-to-Dewar Isomerisation in BN-Doped Naphthalene: Mechanistic Implications for Molecular Solar Thermal Storage}} \\%Article title goes here instead of the text "This is the title"
\vspace{0.3cm} & \vspace{0.3cm} \\

 & \noindent\large{Michael Bühler,\textit{$^{a}$} Merle I. S. Röhr$^{\ast}$\textit{$^{a,b}$}} \\%Author names go here instead of "Full name", etc.

\includegraphics{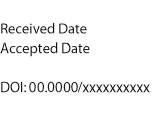} & \noindent\normalsize{The planar to Dewar valence isomerisation of 4a,8a-azaboranaphthalene (\bnNaph), a $\pi$ extended \ce{BN}-doped analogue of azaborine, is investigated to evaluate how \ce{BN} incorporation reshapes the minimum energy pathway on the ground state. This process is, for example, relevant in the context of molecular solar thermal (MOST) energy storage, where absorbed sunlight is converted into chemical energy through reversible photoisomerisation. Structures and vertical excitations were computed using DFT and TD-DFT, minimum energy pathways were mapped with nudged elastic band (NEB) calculations, and pathway energetics were refined with state averaged XMS-CASPT2. In addition, azaborine was examined as a comparison system, with particular emphasis on whether substituents at nitrogen and boron promote Dewar formation. The effect of \ce{BN} doping on the system was analysed in detail. Compared with the carbon analogue, the conversion pathway becomes asymmetric with a metastable intermediate stabilized by a transient boron to carbon contact. The transition structure closely resembles an S$_0$/S$_1$ conical intersection, which is consistent with a vibrationally activated nonradiative funnel. For tuning MOST properties, screening of single substituents across the whole molecule reveals predominantly red shifted S$_1$ energies together with increased oscillator strengths and indicates that appropriate substitution can improve Dewar formation in azaborine derivatives.} \\%The abstrast goes here instead of the text "The abstract should be..."

\end{tabular}

 \end{@twocolumnfalse} \vspace{0.6cm}

  ]
%%%END OF TITLE, AUTHORS AND ABSTRACT%%%

%%%FONT SETUP - please do not change any commands within this section
\renewcommand*\rmdefault{bch}\normalfont\upshape
\rmfamily
\section*{}
\vspace{-1cm}

%%%FOOTNOTES%%%

\footnotetext{\textit{$^{a}$~Institute of Physical and Theoretical Chemistry, University of Würzburg, Emil-Fischer-Str. 42, 97074 Würzburg, Germany \\ \textit{$^b$} Center for Nanosystems Chemistry, Universität Würzburg, Theodor-Boveri-Weg, 97074 Würzburg Germany. E-mail: merle.roehr@uni-wuerzburg.de}}

%Please use \dag to cite the ESI in the main text of the article.
%If you article does not have ESI please remove the the \dag symbol from the title and the footnotetext below.
\footnotetext{\dag~Supplementary Information (SI) available. See DOI: 10.1039/cXCP00000x/}
%additional addresses can be cited as above using the lower-case letters, c, d, e... If all authors are from the same address, no letter is required

%%%END OF FOOTNOTES%%%

%%%MAIN TEXT%%%%

\section{Introduction}

Molecular solar thermal (MOST) energy storage \cite{sunApplicationsPhotoswitchesStorage2019,xuMolecularSolarThermal2025, bieblHighEnergyDensity2025} encompasses a family of concepts in which absorbed sunlight is converted into chemical energy by a reversible photochemical transformation, stored in a metastable photoisomer, and released as heat on demand through thermal or catalytic back conversion.\cite{gimenez-gomezPhotochemicalOverviewMolecular2022} Distinct MOST implementations can be differentiated by the underlying switching motif and reaction class. These include $E/Z$ isomerisations of stilbene and azo compounds,\cite{boelkeDesigningMolecularPhotoswitches2019,franzElectrochemicallyTriggeredEnergy2022,qiuVisibleLightActivated2023,kobauriRationalDesignPhotopharmacology2023} photoinduced valence isomerisations such as norbornadiene--quadricyclane,\cite{grayDiarylsubstitutedNorbornadienesRedshifted2014,kuismaComparativeAbInitioStudy2016,jornerUnravelingFactorsLeading2017,orrego-hernandezEngineeringNorbornadieneQuadricyclane2020,hemauerNorbornadieneQuadricyclanePair2024,kochmanSimulatingEnergyCapture2025} dimerisations,\cite{chakrabortySelfactivatedEnergyRelease2024} and photoinduced ring opening processes exemplified by dihydroazulene--vinylheptafulvene.\cite{lindbaekbromanDihydroazuleneControllingPhotochromism2014,schottlerLongTermEnergyStorage2022}

Despite these mechanistic differences, the central performance targets are shared across platforms, namely solar match of the absorption band, spectral separation between the parent and storage isomer to suppress competitive photochemistry, a high quantum yield of photocharging, a sufficiently long storage lifetime, and a high gravimetric energy density.\cite{borjessonEfficiencyLimitMolecular2013,gimenez-gomezStateoftheartChallengesMolecular2024,schatzMolecularSolarThermal2026}

\begin{figure}
    \centering
    \includegraphics[]{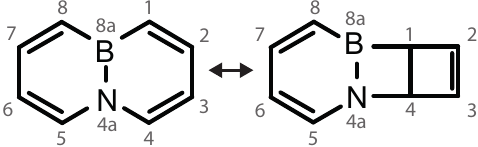}
    \caption{Lewis structures of the planar \bnNaph isomer (left) and the Dewar valence isomer (right). Numbering follows IUPAC conventions and is used throughout the text.}
    \label{fig:LEWIS}
\end{figure}

Prominent reference points are the \ce{BN}-doped \Benz scaffold 1,2-dihydro-1,2-azaborine (\bnBenz)\cite{broughPhotoisomerization12Dihydro12AzaborineMatrix2012,edelDewarIsomer12Dihydro12azaborinines2018,richterFacileEnergyRelease2024} or the DNA-inspired pyrimidone,\cite{nguyenMolecularSolarThermal2026} which have been established as MOST candidates through the selective formation of their Dewar valence isomer and the high energy storage density of the parent motif. Within this molecular class, the rich and long known landscape of \Benz valence isomers becomes accessible, including fulvene, benzvalene, and Dewar-type configurations.\cite{wardVacuumUltravioletPhotolysis1968,johnstoneMatrixcontrolledPhotochemistryBenzene1991,pieriNonadiabaticNanoreactorAutomated2021} Beyond the generation of individual isomers within the BN-doped scaffold, extensive work has also addressed their mutual rearrangements, in particular the conversion of the Dewar isomer into the benzvalene-like structure.\cite{ozakiBNBenzvalene2024,guerraBNDewarBenzeneBNbenzvalene2025,bieblSwitchingShapesReversible2026} 

For the \bnBenz molecule, the photoinduced conversion from the planar form to the Dewar isomer is supported by a consistent picture emerging from experiment and high-level electronic-structure calculations, including CCSD(T) and (XMS-)CASPT2 treatments.\cite{bettingerRingOpening2aza3borabicyclo220hex5ene2013,suMechanisticInvestigationsPhotoisomerization2013,ozakiBoronNitrogenContainingBenzeneValence2024,kimTheoreticalInvestigationReaction2015} Upon excitation into S$_1$, the molecule can access a prefulvene-like intermediate, which subsequently evolves toward the Dewar product. This productive channel competes with nonproductive deactivation pathways, most prominently internal conversion through a conical intersection that returns the system directly to the planar ground-state minimum.\cite{jeongUltrafastPhotoisomerizationMechanism2023,m.arpaPhotochemicalFormationElusive2024} 
The overall isomerisation efficiency can be substantially improved by targeted substitution of the azaborine scaffold, with a 1-(tert-butyldimethylsilyl) group at nitrogen and either a mesityl substituent or chlorine at boron, as exemplified by TBS/Mes-azaborine and TBS/Cl-azaborine, respectively.\cite{marwitzHybridOrganicInorganic2009,edelDewarIsomer12Dihydro12azaborinines2018,ozakiPositionalIsomerization12Azaborine2026}
Furthermore, to reduce computational cost, a chemically simplified model in which TBS was replaced by a silyl group was investigated theoretically for the chloro-substituted case (SiH$_3$/Cl-azaborine).\cite{m.arpaPhotochemicalFormationElusive2024}
This improvement has been rationalised by a reshaping of the ground-state landscape that eliminates or destabilises the intermediate and thereby reduces kinetic branching into unproductive channels.\cite{m.arpaPhotochemicalFormationElusive2024}

Related Dewar-type photoisomerisations are likewise documented for larger polycyclic aromatic systems such as \Naph and anthracene, underscoring the generality of these valence rearrangements across extended aromatics.\cite{guestenDeactivationFluorescentState1980,mikiSynthesisPhotoreaction1234Tetratbutylnaphthalene1992,latajkaCNDO2Molecular1981,chakrabortyCurvedAnthracenesVisiblelight2025}
Within this framework, the substitution of \ce{CC} units by \ce{BN} in aromatic scaffolds has emerged as a widely explored strategy to expand the chemical space of organic molecules.\cite{dewar624NewHeteroaromatic1958,dewarNewHeteroaromaticCompounds1968,kawaguchiMaterialsBasedGraphite1997,liuBNCCHow2008,bosdetBNCCSubstitute2009,campbellRecentAdvancesAzaborine2012,morganEfficientSyntheticMethods2016,ishibashiBNTetraceneExtending2017,giustraStateArtAzaborine2018,mcconnellLatestageFunctionalizationBNheterocycles2019,rulliPropenolysisEnyneMetathesis2024,rulliCatellaniInspiredBNAromaticExpansion2026} 

Beyond MOST, \ce{BN}-substituted aromatics have also been proposed as candidates for singlet-fission materials,\cite{zengSeekingSmallMolecules2014,pinheiroSystematicAnalysisExcitonic2020,singhEnergeticsOptimalMolecular2021,zhangElectronDelocalizationLowlying2023} underscoring that \ce{BN} incorporation can tune excited-state energetics in functionally relevant ways. Consistent with this broader potential, recent years have seen growing efforts to elucidate the spectroscopy and photochemistry of \ce{BN}-substituted aromatic molecules across different size regimes and environments.\cite{holzmeierPhotoionizationPyrolysis14Azaborinine2014,snyderBNDopingPhotochemistry2017,snyderExcitedStateDeactivationPathways2017,nazariUltrafastDynamicsPolycyclic2019,appiariusBNSubstitutionDithienylpyrenesPrevents2022} 
These advances motivate transferring MOST-oriented design concepts from \ce{BN}-doped \Benz, where Dewar formation is well characterised, to larger \ce{BN}-doped polycyclic aromatic hydrocarbons with one additional fused ring. Such scaffold extension is expected to bathochromically shift the relevant absorption band toward the near-UV. In these \ce{BN}-doped naphthalene molecules, different doping patterns have already been shown to tune the absorption toward wavelengths beyond \SI{300}{nm}.\cite{liuLeastStableIsomer2017} Notably, \bnNaph (see Fig. \ref{fig:LEWIS}) absorbs in this regime and exhibits a $0$--$0$ transition at \SI{295.6}{\nm} (\SI{4.19}{\eV}) in the UVA range, making it a suitable model to assess how \ce{BN} incorporation and $\pi$-extension jointly shape the photophysics of potential MOST chromophores.

The \ce{BN} insertion at the 4a,8a bridgehead positions preserves the overall naphthalene topology while introducing a polarised \ce{BN} unit into the ring framework. In a previous paper,  we theoretically investigated the excited state landscape of isolated \bnNaph, guided by picosecond time resolved spectroscopy measurements in a supersonic free jet.\cite{sturmImpactIsoelectronicSubstitution2024}
The mechanism elucidated for isolated \bnNaph suggests that the doping influences the topology of the excited state potential energy surfaces and thus redirects deactivation and photoisomerisation pathways.
The calculations (SA-XMS-CASPT2 level) indicated that excitation into S$_1$ at the origin, i.e., without excess vibrational energy, does not directly open an efficient nonradiative decay route, consistent with an excited state lifetime on the nanosecond timescale. In contrast, upon vibrational excitation, nonradiative deactivation becomes possible. 
This switch in accessibility is traced to a conical intersection that lies beyond reach at the S$_1$ origin but becomes available upon deposition of additional vibrational energy, and is associated with an out-of-plane distortion of the carbon atom adjacent to nitrogen. 
However, in contrast to the \ce{BN}-doped system, the intersection in \Naph is energetically much less accessible and lies substantially higher relative to the relevant conversion pathway, making it less likely to govern the relaxation dynamics under comparable excitation conditions. 
Experimentally, \Naph is known to exhibit a comparatively long-lived S\(_1\) state and to relax predominantly via radiative decay, which is consistent with the present mechanistic picture.\cite{zgierskiCommentsVibronicIntensities1977,robeyPrioriCalculationsVibronic1977,negriVibronicStructureEmission1996} The relevant conical intersection in the \Naph molecule is located at substantially higher energy and therefore cannot be efficiently reached from the initially populated excited-state region, consistent with the distinct conversion pathway and the absence of an analogous low-energy funnel. A detailed discussion of these energetic differences and their consequences for the photophysical behaviour is given in Sturm et al.\cite{sturmImpactIsoelectronicSubstitution2024}

\begin{figure}
    \centering
    \includegraphics[]{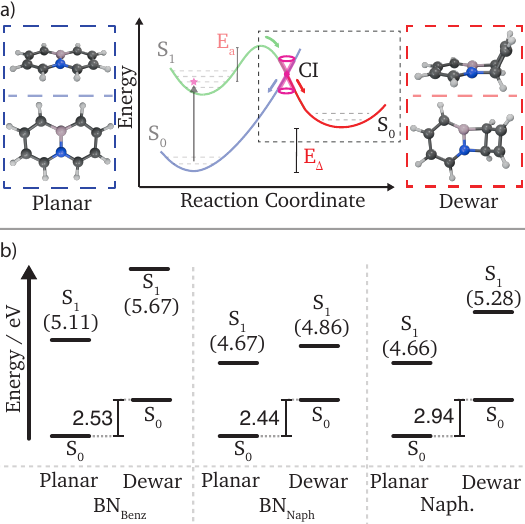}
    \caption{a) Potential energy profile along the pathway connecting the planar and Dewar isomers, showing the S$_0$ (blue) and S$_1$ (green) surfaces, the approach to the conical intersection (CI), and subsequent relaxation to the product minimum. The pathway adopted in this work, highlighted in red, follows the route toward the Dewar minimum. The remaining segments correspond to the previously reported pathway from Ref.~\cite{sturmImpactIsoelectronicSubstitution2024}. $E_\mathrm{a}$ denotes the barrier on S$_1$ and $E_\Delta$ the relative energy offset between the ground state minima. b) Comparison of the $E_\Delta$ values between the planar and Dewar minima on the S$_0$ surface for \bnBenz, \bnNaph and \Naph. The values given in parentheses correspond to the S$_1$ excitation energies of the respective optimized structures. All energies were computed at the $\omega$B97X-D3/aug-cc-pVDZ level of theory.}
    \label{fig:Pathway_extended}
\end{figure}

A central question for assessing the relevance of \bnNaph in the context of MOST is whether deactivation pathways analogous to those of \bnBenz, including access to alternative photoproducts, are also available in this system and which conclusions can be drawn from their comparison (see Fig.~\ref{fig:Pathway_extended}). To address this question, we first examine the effect of \ce{BN} incorporation on the electronic structure, then analyse the conversion pathway of \bnBenz as a reference, and finally relate these findings to \bnNaph. This comparison allows us to identify which mechanistic features remain transferable upon $\pi$-extension and \ce{BN} incorporation, whether substitution can steer the reactivity in analogy to \bnBenz, and whether the resulting photochemical behavior fulfills the key criteria for MOST applications.
\section{Methods}

Quantum chemical calculations aimed at elucidating the spectra within the framework of (time-dependent) Density Functional Theory (TD-DFT). The calculations employed the $\mathrm{\omega}$B97XD functional\cite{linLongRangeCorrectedHybrid2013} along with the augmented correlation consistent polarised double zeta (aug-cc-pVDZ) basis set.\cite{kendallElectronAffinitiesFirstrow1992} All DFT and TD-DFT computations were executed using the Orca 6.0.1 quantum chemical software package.\cite{neeseSoftwareUpdateORCA2025} The molecular structures were optimised, and Hessian matrices were computed for all ground (S$_0$) and the transition states. Hirshfeld charges were evaluated using the \textsc{Multiwfn} program.\cite{luComprehensiveElectronWavefunction2024}

To obtain minimum-energy pathways, we interpolated between the planar and Dewar structures using the climbing-image nudged elastic band (CI-NEB) method in the implementation available in Orca, followed by transition-state optimisation.\cite{henkelmanClimbingImageNudged2000,asgeirssonNudgedElasticBand2021} The initial pathway was generated at the GFN2-xTB level with 10 images.\cite{bannwarthGFN2xTBAnAccurateBroadly2019}

To validate the minimum-energy path obtained with DFT, for \bnNaph additionally also the SA-XMS-CASPT2 method was employed as implemented in \textsc{BAGEL}\cite{shiozakiBAGELBrilliantlyAdvanced2018}, using an active space of six electrons in six orbitals and the aug-cc-pVDZ basis set. To avoid possible intruder states in the XMS-CASPT2 calculations, the real level shift of 0.5 Hartree is employed.
The initial pathway was generated with \texttt{geodesic\_interpolate.py},\cite{zhuGeodesicInterpolationReaction2019} inserting 12 intermediate images between the optimised planar minimum and the Dewar endpoint. The pathway was subsequently refined using a custom NEB implementation based on the Henkelman--J\'onsson formalism, employing a parallel BFGS optimisation scheme.\cite{henkelmanImprovedTangentEstimate2000}  Finally, additional structures were inserted between the optimised NEB images using \texttt{geodesic\_interpolate.py}, and single-point calculations were carried out to obtain a higher-resolution energy profile along the transformation coordinate.
Nucleus-independent chemical shifts (NICS) were computed with \texttt{pyAroma\,5},\cite{wangPyAromaIntuitiveGraphical2024} interfaced to Gaussian\,16.\cite{frischetal.Gaussian16RevisionA032016} The same density functional and basis set as in the geometry optimisations were employed.
\section{Results and discussion}

\subsection{Impact of \ce{BN} doping on the planar and Dewar structure}

\begin{figure}[t]
    \centering
    \includegraphics[]{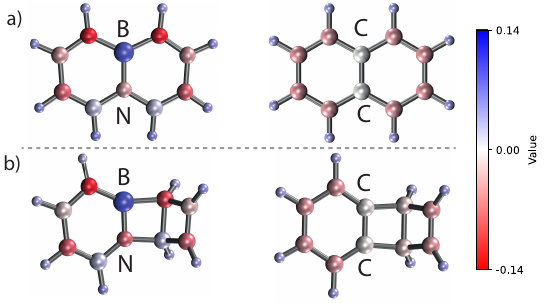}
    \caption{Hirshfeld charges for the planar (top) and the Dewar (bottom) form. Partial atomic charges are mapped onto the molecular surface and shown by the color scale, with red indicating electron poor regions and blue indicating electron rich regions.}
    \label{fig:hirshfeld}
\end{figure}

Although the bridging carbon atoms are replaced isoelectronically by a boron-nitrogen unit, the polar \ce{B-N} bond still induces a pronounced shift of electron density across the molecule. In the \ce{BN}-doped systems, boron carries a substantially more negative partial charge and nitrogen becomes correspondingly more positive, whereas the analogous carbon sites in \Naph remain close to charge neutral. This polarisation is not limited to the heteroatoms themselves, since the directly bound carbon atoms also show marked changes in their Hirshfeld charges relative to the \ce{CC} analogue. The same qualitative charge pattern is obtained for the Dewar isomers (see Fig. \ref{fig:hirshfeld}).

\begin{figure}[t]
    \centering
*     \includegraphics[]{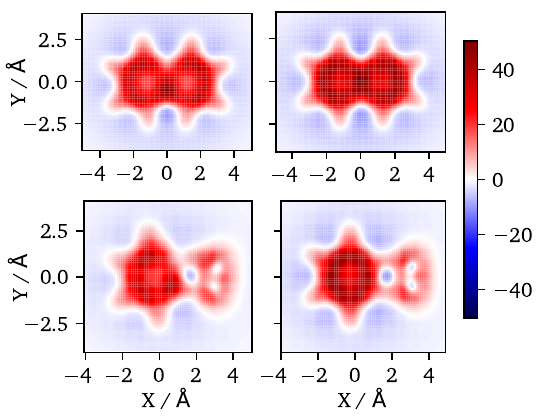}
    \caption{Two dimensional NICS map evaluated in a plane parallel to the $\pi$ framework and positioned \SI{0.8}{\angstrom} above the ring system. All geometries were aligned such that the molecular $\pi$ scaffold lies in the XY plane, and the NICS values were sampled on an XY grid in this plane. Similar to Fig.~\ref{fig:hirshfeld}, the left column displays the \ce{BN}-doped species, whereas the right column shows the corresponding all-carbon analogues. The top row corresponds to the planar isomers, and the bottom row to the Dewar valence isomers. The color scale reports the magnitude and sign of the NICS response across the section plane.}
    \label{fig:NICS_xy}
\end{figure}

This effect is also reflected in the bond lengths. The \ce{C-N} bonds are shortened whereas the \ce{B-C} bonds are elongated, which distorts the framework and shifts the opposing \ce{CH} groups from an almost parallel arrangement in \Naph to a weakly V-shaped geometry. The \ce{C4-N-C5} and \ce{C1-B-C8} bond angles are \SI{118.5}{\degree} and \SI{128.8}{\degree}, respectively, while in \Naph, the corresponding angles are \SI{122.0}{\degree} on both sides. Further details on the planar structure are provided by Sturm et al.~\cite{sturmImpactIsoelectronicSubstitution2024}

During formation of the Dewar isomer, the six-membered ring is converted into a  bicyclo[2.2.0] framework, accompanied by formation of a new \ce{C1-C4} bond. Accordingly, the distance between the reacting carbon atoms contracts from \SI{2.86}{\angstrom} in planar \bnNaph to \SI{1.60}{\angstrom} in the Dewar form. An analogous contraction is found for \Naph, from \SI{2.80}{\angstrom} to \SI{1.58}{\angstrom}. In addition, the bond lengths involving the bridging atoms shorten upon Dewar formation, decreasing from \SI{1.47}{\angstrom} to \SI{1.43}{\angstrom} for \bnNaph and from \SI{1.42}{\angstrom} to \SI{1.39}{\angstrom} for \Naph.

Concomitantly, a pronounced twist of the remaining six-membered ring is observed in the \ce{BN}-doped compound relative to the \ce{CC} analogue (see Fig.~\ref{fig:hirshfeld}). The \ce{C4-N-C5} angle amounts to \SI{137.9}{\degree}, whereas the corresponding \ce{C1-B-C8} angle is substantially larger (\SI{153.4}{\degree}); in the \Naph reference, both angles are identical and equal to \SI{144.1}{\degree}. 

The same qualitative trends are also observed for the smaller \bnBenz unit compared with \Benz when going from the planar to the Dewar form. In \bnBenz, the \ce{C-B-H} angle is \SI{140.0}{\degree} in contrast to the \ce{C-N-H} angle of \SI{126.6}{\degree}. For \Benz, the corresponding angles are \SI{120.0}{\degree} on both sides.

Heteroatom doping affects both the molecular geometry and the electronic properties of the $\pi$ system. To assess the resulting changes in $\pi$ delocalisation, NICS$_\perp$ values were calculated perpendicular above and below the ring centres assigned in Fig.~S1 for the doped and undoped systems in both the planar and Dewar geometries (see Tab.~\ref{tab:points_form_compact_spaced}).\cite{stojanovicMonoBNsubstitutedAnalogues2018} The comparison shows a similar redistribution of aromatic character in both systems upon Dewar formation. The first six-membered ring becomes more aromatic, as indicated by more negative NICS values, whereas the two newly formed four-membered rings become distinctly antiaromatic and exhibit strongly positive values, consistent with paratropic ring currents expected from Hückel's rule.

Cross-sections through the NICS maps are shown in Fig.~
\ref{fig:NICS_xy} and visualise the spatial distribution of shielding and deshielding regions. Negative, diatropic regions above the six-membered rings support their aromatic character, while positive, paratropic regions around the four-membered rings indicate the antiaromatic contribution introduced by Dewar formation. In the \bnNaph, the heteroatom substitution breaks the symmetry of \Naph and produces a stronger paratropic contribution on the nitrogen side than on the boron side.

In aromatic systems, incorporation of a \ce{BN} unit at the bridged position generally perturbs the $\pi$ electron distribution and often reduces the overall degree of aromaticity, as reported previously.\cite{zhangElectronDelocalizationLowlying2023} Accordingly, the energy gap between the planar and Dewar structures differs for \Naph and \bnNaph. For \bnNaph, this gap amounts to \SI{2.44}{\electronvolt}, whereas for \Naph it is larger by about \SI{0.50}{\electronvolt} (see Fig. \ref{fig:Pathway_extended}b). These sizeable energy differences reflect the pronounced reorganisation of the electronic structure, particularly the change in $\pi$-electron delocalisation, upon planar to Dewar rearrangement.

Overall, these results show that \ce{BN} doping introduces a pronounced electronic and geometric asymmetry that redistributes charge density, perturbs aromatic stabilisation, and thereby lowers the energetic separation between the planar and Dewar forms.

% frei einstellbarer Shift pro B-Label (z.B. -0.3em oder 0.2em)
\newcommand{\Bshift}[2]{\makebox[0pt][c]{\hspace{#1}#2}}

\begin{table}[ht]
\centering
\caption{Perpendicular NICS$_\perp$ point values (B) of the different ring systems (for notation cf. Fig. S1).}
\label{tab:points_form_compact_spaced}
\setlength{\tabcolsep}{3pt}
\renewcommand{\arraystretch}{1.20}
\small
\begin{tabular}{
c c
@{\hspace{14pt}}
S[table-format=-2.2,table-number-alignment=center] S[table-format=-2.2,table-number-alignment=center]
@{\hspace{14pt}}
S[table-format=-2.2,table-number-alignment=center] S[table-format=-2.2,table-number-alignment=center]
@{\hspace{14pt}}
S[table-format=-2.2,table-number-alignment=center] S[table-format=-2.2,table-number-alignment=center]
}
\toprule
\multirow{2}{*}{Form} & \multirow{2}{*}{Mat.} &
\multicolumn{2}{c}{\Bshift{-0.4em}{Ring 1}} &
\multicolumn{2}{c}{\Bshift{-0.4em}{Ring 2}} &
\multicolumn{2}{c}{\Bshift{0.8em}{Ring 3}} \\
\cmidrule(l{0.2em}r{1.5em}){3-4}
\cmidrule(l{1.2em}r{1.5em}){5-6}
\cmidrule(l{0.8em}r{0.8em}){7-8}
 &  &
\multicolumn{1}{c}{\Bshift{0.6em}{B1}} &
\multicolumn{1}{c}{\Bshift{-0.45em}{B2}} &
\multicolumn{1}{c}{\Bshift{0.7em}{B3}} &
\multicolumn{1}{c}{\Bshift{-0.45em}{B4}} &
\multicolumn{1}{c}{\Bshift{1.0em}{B5}} &
\multicolumn{1}{c}{\Bshift{1.2em}{B6}} \\
\midrule
\multirow{2}{*}{Dewar}
  & \ce{BN} & -19.98 & -20.31 &   0.84 &   1.74 &   1.50 &   0.44 \\
  & \ce{CC} & -28.64 & -27.70 &   3.15 &   1.56 &   1.03 &   0.03 \\
\multirow{2}{*}{Planar}
  & \ce{BN} & -20.48 & -20.48 & -20.48 & -20.48 & \multicolumn{1}{r}{\Bshift{-1.6em}{--}} & \multicolumn{1}{r}{\Bshift{-1.6em}{--}} \\
  & \ce{CC} & -28.91 & -28.91 & -28.91 & -28.91 & \multicolumn{1}{r}{\Bshift{-1.6em}{--}} & \multicolumn{1}{r}{\Bshift{-1.6em}{--}} \\
\bottomrule
\end{tabular}
\end{table}

\subsection{Isomerisation pathway of \bnBenz}

 To place the planar-Dewar transformation of \bnNaph into a broader context, the ground-state isomerisation pathways of the established reference system \bnBenz and its substituted derivatives, TBS/Mes-azaborine and \ce{SiH3}/\ce{Cl}-azaborine, were first investigated. To this end, minimum-energy path calculations were carried out for all three systems using the nudged elastic band method. The resulting NEB profiles complement the discrete transition-state information available from the literature.\cite{m.arpaPhotochemicalFormationElusive2024}
To establish a consistent baseline for comparison between \bnNaph and \Naph, an additional reference calculation was carried out for \Benz. In order to maintain a uniform and chemically intuitive comparison throughout this work, the atom-labeling scheme for \bnNaph follows the notation introduced in Fig.~\ref{fig:LEWIS}, rather than the corresponding IUPAC nomenclature; the complete labeling convention is shown in Fig.~S2.
 
\begin{figure}[h]
    \centering
    \includegraphics[]{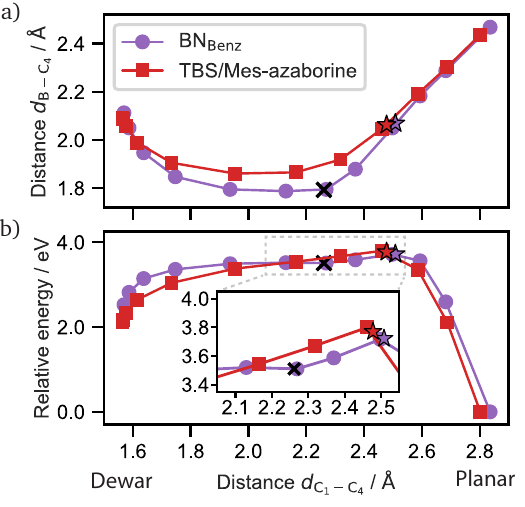}
    \caption{Nudged elastic band pathway connecting the planar and Dewar forms for two \bnBenz systems, the 1,2-dihydro parent compound and the disubstituted derivative bearing a TBS substituent at nitrogen and a Mes substituent at boron. Both panels are plotted along the transformation coordinate defined by the \(d_{\ce{C1-C4}}\) distance on the x-axis. a) \(\ce{B-C4}\) distance, denoted as \(d_{\ce{B-C4}}\), along the transformation. b) Corresponding relative energy profile along the same transformation coordinate, with an inset magnifying. The \bnBenz profile exhibits a shallow local minimum in this region, whereas this minimum disappears upon substitution with TBS and Mes, resulting in a monotonic energy decrease and indicating destabilisation of the intermediate along the ground-state pathway. The positions of the corresponding optimised transition states are marked by star symbols, while the identified optimised local minimum is indicated by a black cross.}
    \label{fig:Benz_Rest}
\end{figure}

Overall, the structural evolution along the conversion pathway is similar for systems with and without substituents at boron and nitrogen. The distance between boron and the carbon atom adjacent to nitrogen, \ce{C4}, \(d_{\ce{B-C4}}\), provides a useful descriptor of this process (see Fig.~\ref{fig:Benz_Rest}a). 
During the initial step of the planar-to-Dewar transformation, pyramidalisation of \ce{C4} leads to a prefulvene-like arrangement, while \(d_{\ce{B-C4}}\) decreases continuously from \SI{2.47}{\angstrom} to \SI{1.79}{\angstrom}. Along this coordinate, \bnBenz passes through a well-defined transition state (see Fig.~\ref{fig:Benz_Rest}b). Beyond this point, the methyl group adjacent to boron rotates out of plane and the boron atom moves downward. Concurrently, the new \ce{C1-C4} bond forms, while the \ce{B-C4} bond elongates markedly. In the final stage, the substituents at boron and nitrogen reorient to align with the plane of the newly formed four-membered ring. 

\begin{figure*}[t]
    \centering
    \includegraphics[]{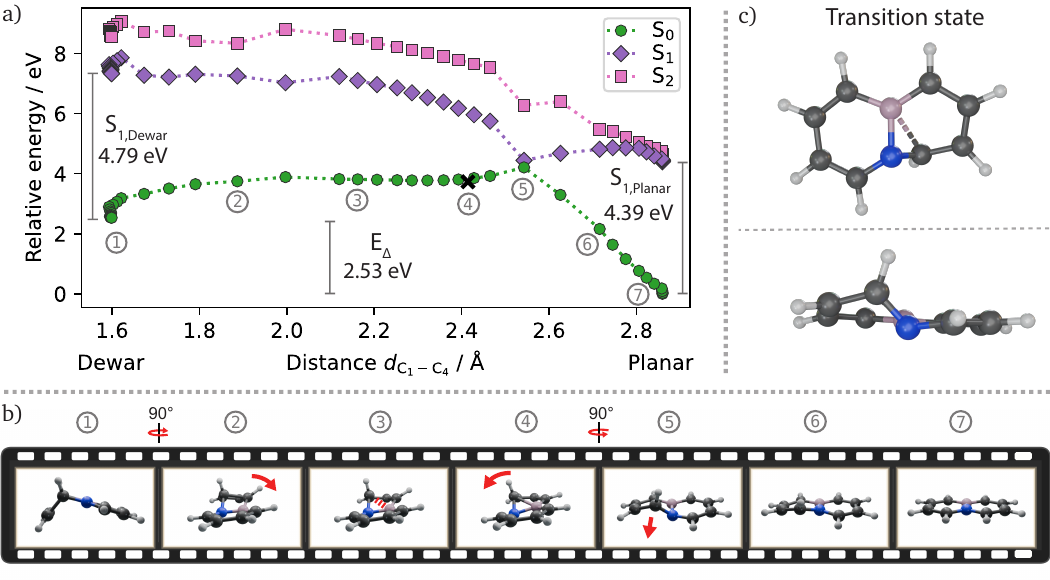}
    \caption{a) Converged nudged elastic band minimum energy path with additional linearly interpolated images between adjacent structures. Single point energies were computed at the XMS-CASPT2 level using an active space of (6,6) with state averaging over four electronic states. Along the transformation coordinate, the left side corresponds to the planar isomer and the right side to the Dewar isomer in the energy profile. Notably, the S$_0$ and S$_1$ energy curves approach each other in the transition state region, indicating a reduced S$_0$–S$_1$ gap at this geometry. The optimised local minimum, marked by a black cross, displays pronounced perfulvene-like character associated with formation of a new B–N bond. b) Principal structural snapshots of the isomerisation pathway, marked in the energy profile. c) NEB transition-state geometry corresponding to Image~5 in panel b, shown in bird's-eye view (top) and front view (bottom), with the forming boron--carbon bond indicated by dotted lines.}
    \label{fig:NEB_PATH_TS_STATE}
\end{figure*}

This structural reorganisation is sensitive to the substitution pattern at boron and nitrogen. In particular, the TBS/Mes-azaborine is known to form the Dewar isomer markedly more efficiently. \cite{edelDewarIsomer12Dihydro12azaborinines2018,m.arpaPhotochemicalFormationElusive2024} One mechanistic factor contributing to this behaviour is the absence of the local minimum along the conversion pathway. This difference is clearly reflected in the evolution of the newly formed \ce{B-C4} bond between the boron and carbon atoms. In \bnBenz, a significantly more stable intermediate region is formed, characterised by a shorter \ce{B-C4} bond length of \SI{1.78}{\angstrom}, which persists over a larger part of the pathway. In the substituted derivatives, by contrast, the newly formed \ce{B-C4} bond remains longer, amounting to \SI{1.86}{\angstrom} in the TBS/Mes-azaborine and \SI{1.90}{\angstrom} in \ce{SiH3}/\ce{Cl}-azaborine, while the energy decreases directly without the appearance of a local minimum (see Figs.~\ref{fig:Benz_Rest} and S3). Thus, although the overall sequence of structural changes along the pathway remains the same, the substituted system follows a more direct isomerisation pathway with a less pronounced \ce{B-C} interaction. A visual representation of the isomerisation pathway is provided in Animations~1 and 2.

For all three \bnBenz-like molecules, further geometry optimizations were carried out using as starting structure the geometry located at the position of the distinct local minimum between the planar and Dewar forms. For \bnBenz, this structure converged to a stationary point without imaginary frequencies and therefore represents a metastable intermediate, separated by a barrier of \SI{0.21}{eV} from the planar structure and by a small barrier of \SI{0.01}{eV} from the Dewar structure. In contrast, for the substituted derivatives, the corresponding optimizations proceed directly toward the Dewar form, which is obtained as the final optimization product. This indicates that substitution destabilizes the intermediate \ce{B-C} interaction and thus the prefulvene-like transition region. The resulting energy profiles are shown in Figs.~\ref{fig:Benz_Rest} and S3. Overall, replacement of hydrogen by substituents leads to a clear change in the conversion profile. This is consistent with the qualitative picture reported by Arpa et al.\cite{m.arpaPhotochemicalFormationElusive2024}

In contrast, the \Benz shows a different behaviour. In the first step, one of the \ce{CH} unit puckers out of the plane. However, it remains essentially symmetric with respect to the ring framework and does not preferentially approach either side. Consequently, no directional shortening towards one partner, as observed for the approach towards boron in the \ce{BN} system, is found. At the transition state, the hydrogen atom of the upward oriented \ce{CH} group points towards the opposite side of the ring.
In the second step, the other side of the framework bends upward, the \ce{CH} group rotates outward again, and the \ce{C1-C4} bond is formed (see Animation~3). Importantly, substitution modifies this qualitative conversion pathway, as demonstrated for 1-silyl-2-chlorobenzene. The resulting pathway can be viewed as intermediate between those of \Benz and \bnBenz, proceeding through a \Benz like rearrangement but with clear asymmetry and a pervulene-like transition state structure (see Fig. S4). The substituents therefore favour an alternative pathway in which formation of a transient bond becomes possible. In \bnBenz, this bond formation significantly stabilizes the distorted structure and leads to the emergence of a corresponding stationary point along the transformation coordinate.

Overall, these results indicate that the qualitative isomerisation mechanism is preserved across the \ce{BN} and \ce{CC} reference systems, even though the corresponding transition states are not identical. The boron atom and the substitution pattern mainly control the degree of stabilisation of intermediate structures and thereby influence how directly the system converts into the Dewar form. Furthermore, in contrast to \Benz, where the transformation proceeds without directional preference, the \ce{BN} systems display a distinct asymmetry along the pathway.

\subsection{Isomerisation pathway of \bnNaph}

Viewed from the Dewar form back to the planar structure, the isomerisation pathway of \bnNaph closely resembles that found for \bnBenz (see Fig.~\ref{fig:NEB_PATH_TS_STATE}a). The corresponding image sequence is shown in Fig.~\ref{fig:NEB_PATH_TS_STATE}b and Animation~4. Starting from the Dewar configuration (Image~1), the \ce{C1-C4} bond elongates as \ce{C1} moves toward the molecular plane (Image~2). At the same time, $d_{\ce{B-C4}}$ decreases, reflecting the approach of boron toward \ce{C4} and the formation of a transient \ce{B-C4} contact (Image~3). This rearrangement leads to a local minimum on the potential energy surface (Image~4), which is stabilised by approximately \SI{0.16}{\eV} with respect to the Dewar structure and by \SI{0.48}{\eV} with respect to the planar structure. This intermediate is a true minimum, since it can be fully optimised. Further progression involves cleavage of the transient \ce{B-C4} contact and folding of \ce{C4} back toward the molecular plane, giving rise to the transition state (Image~5). Beyond this point, \ce{C1} continues its motion into the plane (Image~6), and the conjugated planar system is restored (Image~7). Thus, the decrease and subsequent increase of $d_{\ce{B-C4}}$ provide a clear structural marker of the \ce{BN}-doped isomerisation pathway.

\begin{figure}[t]
    \centering
    \includegraphics[]{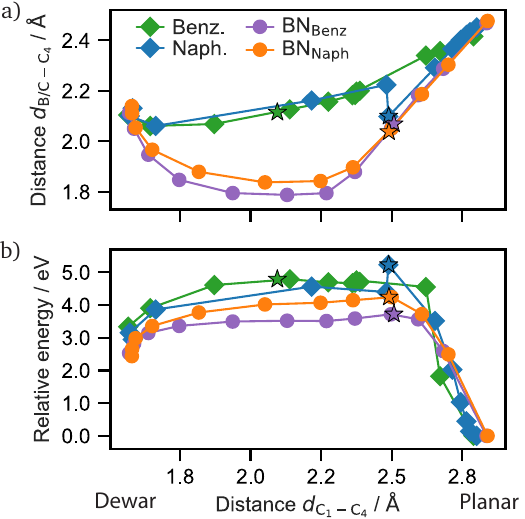}
    \caption{Comparison of converged nudged elastic band isomerisation pathways for \bnBenz and \bnNaph and their all-carbon analogues. Both panels are plotted along the conversion pathway defined by the \(d_{\ce{C1-C4}}\) distance on the x-axis. a) Distance between the boron atom and the carbon atom at position \ce{C4}, denoted as \(d_{\ce{B-C_4}}\) for the \ce{BN}-doped systems, together with the corresponding \(d_{\ce{C-C_4}}\) distance for the CC analogues. b) Relative energy profile along the same transformation coordinate. The positions of the corresponding optimised transition states are marked by star symbols.}
    \label{fig:Dist_Benz_Naph}
\end{figure}

In contrast, \Naph does not show a comparable decrease followed by re-elongation in the analogous distance coordinate. Instead, the separation decreases monotonically along the pathway, consistent with a more continuous structural reorganisation without an explicitly stabilised intermediate geometry. The same qualitative difference between \ce{BN}-doped and \ce{CC} pathways is also observed \Benz/\bnBenz systems as shown in Fig.~\ref{fig:Dist_Benz_Naph}.

\begin{figure}[h]
    \centering
    \includegraphics[width=0.6\linewidth]{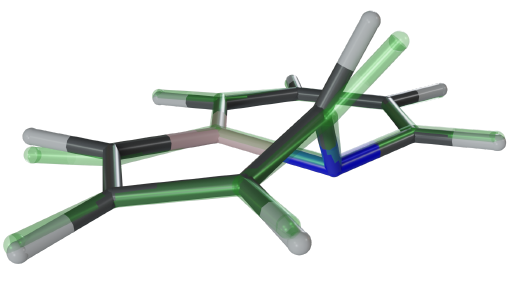}
    \caption{Structural comparison between the transition state for the planar to Dewar isomerisation of \bnNaph and the minimum energy conical intersection identified along the nonradiative deactivation pathway, shown in green. The geometries are superimposed to facilitate a direct assessment of their similarity in the out of plane distortion of the ring framework and in the approach at the reactive boron carbon contact. The close correspondence between both stationary points indicates that the deactivation conical intersection is accessed in the vicinity of the isomerisation bottleneck.}
    \label{fig:CompareTS_CI}
\end{figure}

For \bnNaph, the transition state is not only the highest point along the planar-to-Dewar isomerisation pathway, but it also lies in the immediate vicinity of a conical intersection between the ground state and the first excited state. In other words, the transformation bottleneck on S\(_0\) occurs at a geometry where S\(_0\) and S\(_1\) become nearly degenerate, which provides an efficient funnel for nonradiative relaxation.
Structurally, the NEB transition state closely resembles the minimum-energy conical intersection that was identified for \bnNaph as the key deactivation channel accessed after vibrational activation on S\(_1\). The similarity is reflected by an RMSD of only \SI{0.13}{\angstrom} between both structures. The structural overlap, shown in Fig.~\ref{fig:CompareTS_CI}, highlights the common distortion motif, most prominently the characteristic out-of-plane displacement of the \ce{CH} group on the nitrogen side, which accompanies the approach toward the bond-forming region.
In \Naph, a conical intersection with a broadly similar geometric motif can likewise be identified. The transition state located for the isomerisation, however, shows more pronounced deviations from this conical intersection (\SI{0.43}{\angstrom}). Most notably, the hydrogen atom at \ce{C4} is tilted toward \ce{C1}. In both cases, and more strongly than in \bnNaph, the adjacent hydrogen at the \ce{C3} position is displaced to a greater extent, whereas the out of plane folded \ce{CH} unit remains symmetric in the \Naph instead of being bent to one side. For a structural comparison, see Fig.~S5.

Thus, in both \bnNaph and the smaller \bnBenz analogue, transient \ce{B-C} bond formation emerges as a key factor connecting the ground-state isomerisation transition state with the conical intersection associated with excited-state deactivation.

\subsection{MOST criteria of \bnNaph}
In the following, the suitability of the planar-to-Dewar isomerisation of \bnNaph for MOST applications is examined on the basis of spectral separation, excited-state accessibility, storage energy, and thermal stability. 

The computed vertical excitation energies indicate a clear spectral separation between the planar and Dewar \bnNaph isomers. The first singlet excitation ($\text{S}_0 \rightarrow \text{S}_1$) occurs at about \SI{4.39}{\eV} in the planar (parent) form and at \SI{4.79}{\eV} in the Dewar form, an energy gap of \SI{0.40}{\eV} (see Fig. \ref{fig:NEB_PATH_TS_STATE}a). 
This offset satisfies the MOST criterion for spectral selectivity,\cite{j.mullerRationalDesignRedshifted2025} meaning that one isomer can be selectively excited without simultaneous excitation of the other. This separation is relevant for suppressing photochemical back conversion, as excitation of the Dewar form should not efficiently access the excited-state deactivation pathway leading back to the planar structure. 
%The calculated excitation energy is consistent with the experimental assignment for planar \bnNaph, whose S$_1$ state lies in the near-UV region at approximately \SI{4.19}{\eV}.\cite{sturmImpactIsoelectronicSubstitution2024}

Thermal stability and energy density are also promising. The computed barrier for thermal ring-opening of the Dewar isomer back to the planar form is on the order of \SI{1.0}{\eV}. This large activation energy implies a long lifetime of the Dewar form. For reference, the TBS/Mes-azaborine exhibits $E_\text{a} = \SI{1.17}{\eV}$  ($\SI{112.97}{\kJ\per\mol}$; $t_{1/2}\approx 162$ days at \SI{25}{\celsius}),\cite{ozakiBoronNitrogenContainingBenzeneValence2024} so the \bnNaph Dewar isomer should be at least comparably persistent. The stored chemical energy,
\begin{equation}
    \Delta E = \text{S}_{0,\text{Dewar}} - \text{S}_{0,\text{planar}}
\end{equation}
 is about \SI{2.53}{\eV} (\SI{\approx 244}{\kJ\per\mol}). The value is comparable to the \bnBenz reference reported by Bettinger et al.\cite{bettingerRingOpening2aza3borabicyclo220hex5ene2013}, for which a CCSD(T) energy difference of \SI{2.57}{\eV} was calculated, indicating that $\pi$-extension to \bnNaph does not compromise the energy-storage capability.
 
Overall, \bnNaph combines substantial energy storage, pronounced thermal persistence, and clear spectral separation. The low oscillator strength, $f_\text{osc} = 0.009$, of the S$_1$ transition remains less favourable, but it primarily limits the efficiency of photochemical conversion rather than the storage capacity or thermal stability of the system.

\subsection{Chemical modification of \bnNaph}
To examine how the first excitation energy of \bnNaph can be modulated by substitution, a systematic set of singly substituted derivatives was considered.
The experimental absorption energy of the first excited state of \bnNaph is \SI{4.19}{\eV}, corresponding to an absorption in the UVA. The computed oscillator strength for this transition is comparatively small.
Computational screening for the \bnBenz molecule indicates that push--pull substitution can substantially red shift the first excitation energy while retaining the core photochemical features.\cite{j.mullerRationalDesignRedshifted2025} 

To further tune the excitation energy using the same methodology, a series of electron donating and electron withdrawing substituents was introduced at the different carbon positions of \bnNaph\ (see Fig.~\ref{fig:LEWIS}).
The substitution pattern and choice of groups were guided by the Hammett \(\sigma\) scale. The substituents considered were nitro (\ce{-NO2}), cyano (\ce{-CN}), fluoro (\ce{-F}), chloro (\ce{-Cl}), methyl (\ce{-CH3}), ethyl (\ce{-CH2CH3}), methoxy (\ce{-OCH3}), and dimethylamino (\ce{-N(CH3)2}). In each case, only a single substituent was introduced at one position.
\begin{figure}
    \centering
    \includegraphics[]{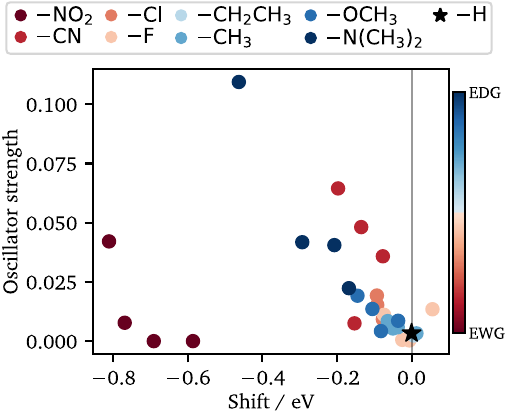}
    \caption{Correlation plot of the computed shift in the vertical excitation energy of the first excited state (S$_1$) versus its oscillator strength for the planar isomer bearing electron donating and electron withdrawing substituents at the \ce{C1}–\ce{C4} positions. Substitutions at \ce{C5}–\ce{C8} correspond to symmetry equivalent carbon sites in the same structures and therefore yield identical values; these points were omitted for clarity. The complete list of all substitution sites is provided in the SI (Tab. S1). The black star denotes the unsubstituted reference structure (R = H).}
    \label{fig:Osc_shift}
\end{figure}

\begin{figure}[t]
    \centering
    \includegraphics[]{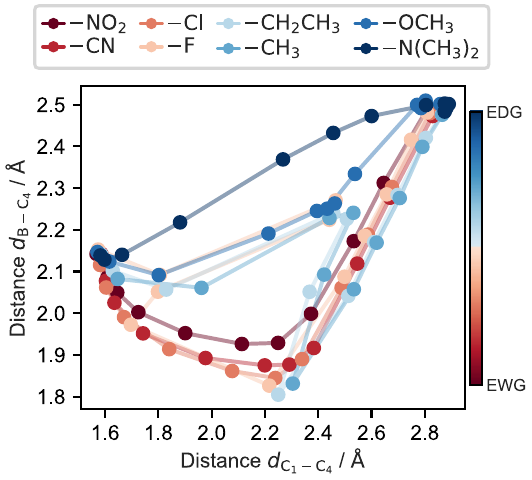}
\caption{Transformation coordinate plot of the conversion pathway between the planar and Dewar forms, showing the distances $d_{\ce{C1-C4}}$ and $d_{\ce{B-C4}}$ for systems substituted at position \ce{C1}. A change in the conversion pathway is observed, evolving from a doubly distorted pathway toward a more $\mathrm{CC_\text{Naph}}$ like character in EDG substituted systems, whereas EWG substituted systems proceed toward $\mathrm{BN_\text{Naph}}$ or $\mathrm{BN_\text{Benz}}$ like structures (compare Fig. \ref{fig:Dist_Benz_Naph}).}

    \label{fig:pos_5}
\end{figure}

For most substitution sites and substituents, a bathochromic shift relative to \bnNaph was obtained, often accompanied by an increased oscillator strength (see Fig.~\ref{fig:Osc_shift}). In particular, dimethylamino substitution lowers the S$_1$ excitation energy by approximately \SI{0.46}{\eV} while substantially enhancing the oscillator strength compared to the parent system. More generally, stronger electron-donating groups (EDGs) or electron-withdrawing groups (EWGs) induce a more pronounced shift of the S$_1$ excitation energy, which can be rationalised by their stronger perturbation of the $\pi$ system of \bnNaph. This effect is particularly pronounced for substituents that extend the conjugated $\pi$ system, as illustrated by the natural transition orbitals of the dimethylamino-substituted derivative, where the $\pi$ system is expanded and the S$_1$ excitation is delocalised over the entire molecular framework (see Fig.~S6). A complete list of excitation energies and oscillator strengths is provided in the Supporting Information (see Tab.~S1).

In addition, for all substituted derivatives the relative energy difference between the planar and Dewar isomers was determined, and the barriers along the corresponding isomerisation pathways were evaluated. Beyond the modulation of the excitation energy, the impact of substitution on the isomerisation pathway itself was analysed in detail. Owing to the symmetry-related substitution sites in \bnNaph, it is possible to disentangle two limiting situations. In one case, the substituent is placed on the ring segment that undergoes the four-membered-ring formation during the planar-to-Dewar rearrangement. In the other case, the substituent is located on the opposite ring and therefore primarily perturbs the electronic structure without being directly involved in the local bond reorganisation. The complete set of NEB calculations for all substituents is provided in the Supporting Information (see Fig.~S7).

For \bnNaph, the characteristic transition-state motif involves transient bonding between boron and the opposing carbon centre that participates in the rearrangement. Substitution can either strengthen this incipient \ce{B-C} bond and stabilise the metastable intermediate, or suppress it such that the intermediate effectively disappears. In \bnNaph, the corresponding \ce{B-C} distance is \SI{1.84}{\angstrom}, while selected substituents shorten it to \SI{1.815}{\angstrom} (Tab.~S2).

Besides the established pathway with a metastable intermediate, two alternative transformation patterns are found for selected substitution sites. The first resembles the behaviour of \Naph and involves out-of-plane motion of a \ce{CH} group, with the attached hydrogen atom rotating toward the ring plane. Here, the \ce{B-C} distance initially decreases as the bond-forming region is approached, but later increases markedly, indicating local geometric reorientation before Dewar formation. The second alternative pathway starts similarly, but shows an approximately linear decrease in the \ce{B-C} separation, which remains above \SI{2}{\angstrom} over much of the transformation. This route is characterised by a more concerted, simultaneous out-of-plane folding on both sides of the ring and is therefore referred to as the double-folding pathway.

Substitution at \ce{C1} has the strongest influence on the isomerisation behaviour and spans all three pathway types. Strong EWGs favour the metastable-intermediate pathway by stabilising the developing \ce{B-C} interaction, whereas increasing electron-donating character shifts the transformation toward alternative routes. Thus, methyl and ethyl substituents follow a swing-by-type mechanism analogous to \Naph, while dimethylamino substitution favours the double-folding pathway (Fig.~\ref{fig:pos_5}). 

For substitution on the less perturbed ring segment, comprising positions 5--8, the pathway remains analogous to that of \bnNaph, and substitution mainly tunes the \ce{B-N} distance and hence intermediate stability. Position 7 is particularly sensitive, with shorter bond distances for EWGs and longer ones for EDGs, consistent with the trends observed for \bnBenz derivatives. All pathways are compiled in the Supporting Information (Fig.~S7 and Tab.~S3).

Inspection of the energetic relationships shows that all substituted systems remain relatively close to \bnNaph. The energy difference between the planar and Dewar structures cannot be substantially increased by substitution and is generally reduced instead. For the \ce{NO2} group, this decrease exceeds \SI{0.4}{eV} relative to \bnNaph (Fig.~S8).

The activation barriers, however, are affected more distinctly. Since these barriers determine the thermal stability of the Dewar isomer, their substituent dependence is particularly relevant. A pronounced example is substitution at position 4, where EWGs increase the barrier, whereas EDGs lower it. For the dimethylamino substituent, the barrier is reduced by approximately \SI{1}{eV} relative to \bnNaph (Fig.~S9).

These observations show that both the nature and position of the substituent influence the planar-Dewar pathway already at the ground-state level, in some cases altering the overall course of the isomerisation. At the same time, substitution can improve the absorption characteristics, particularly the energetic position of the S$_1$ state and its oscillator strength. However, substituents also affect the relative stability of the planar and Dewar forms and therefore the energy-storage properties of the system.

\section{Conclusion}

Previous work has shown that \ce{BN} doping in \bnNaph shifts the energetic position of conical intersections relative to \Naph, making non-adiabatic funnels accessible under excitation conditions where they remain too high in energy for the hydrocarbon analogue. In \bnNaph, the conical intersection is nearly structurally equivalent to the transition state of the planar-to-Dewar ground-state isomerisation (see Fig.~\ref{fig:CompareTS_CI}). Both geometries are dominated by out-of-plane displacement of the carbon atom on the nitrogen side and reorientation of the associated hydrogen toward nitrogen. This correspondence indicates that \ce{BN} substitution renders a transition-state-like geometry electronically accessible, suggesting that the transition state may be reached via nonradiative relaxation from S$_1$ following photoexcitation.

This motivated a closer analysis of how \ce{BN} doping perturbs the aromatic framework of naphthalene. Although formally isoelectronic, the polar \ce{B-N} unit alters local bond metrics and electron density, reshaping the conversion pathway between the planar and Dewar-like forms. The resulting asymmetric, V-shaped framework preorganises the molecule for the planar-to-Dewar transformation and promotes reorganisation toward a fulvene-like intermediate, which can be stabilised by a transient boron-mediated interaction across the ring.

To place the \bnNaph results into context, the established \bnBenz system and the experimentally studied TBS/Mes-substituted \bnBenz derivative were re-examined as reference systems, with particular focus on the conversion pathway. The results agree with previous reports, supporting the validity of the applied approach.\cite{bettingerRingOpening2aza3borabicyclo220hex5ene2013,jeongUltrafastPhotoisomerizationMechanism2023} In  TBS/Mes-azaborine, the peripheral substituents destabilise the boron--carbon-stabilised intermediate, as reflected by the elongated \ce{B-C} distance in this region, and thereby favour a more direct pathway toward the Dewar form. This effect is proposed to contribute to the higher conversion efficiency reported for the substituted \bnBenz system.\cite{m.arpaPhotochemicalFormationElusive2024} By contrast, \Benz lacks the \ce{BN}-induced structural asymmetry and cannot form a comparably stabilised intermediate. Thus, the key difference between the \ce{CC} and \ce{BN} variants arises from a \ce{BN}-driven structural bias that perturbs the $\pi$ system and enables partial carbon--boron bond formation along the transformation coordinate.

With respect to the MOST-related properties, \bnNaph preserves the key design criteria established for \bnBenz, in particular a large energetic separation between the planar and Dewar forms and sufficiently high ground-state barriers to suppress thermal conversion under ambient conditions. However, its main limitation is the low oscillator strength of the S$_1$ transition, which could reduce the efficiency of photoinduced switching. Substitution was therefore explored as a strategy to enhance the S$_1$ absorption while simultaneously tuning the excitation energy and the ground-state isomerisation pathway. Among the substituents considered, electron-donating and electron-withdrawing groups show distinct effects, with a dimethylamino group at carbon position 1 emerging as the most promising modification. It provides a red shift of the S$_1$ excitation by \SI{0.46}{\eV} together with a pronounced increase in oscillator strength, while leaving the overall topology of the ground-state planar-to-Dewar conversion pathway essentially unchanged relative to \bnNaph. The different substitution patterns further show that the rearrangement does not follow a single universal route, but can be grouped into three mechanistic families. Thus, substituent identity and position provide a practical handle to steer the conversion coordinate and to bias the system toward or away from intermediate stabilisation.

In close analogy to \bnBenz reported previously, these findings support the possibility that \ce{BN} doping in naphthalene enables a viable photochemical channel toward the corresponding Dewar valence isomer. While the present work focuses on the ground-state landscape, the photochemical conversion proceeds through an excited state and involves relaxation toward a conical intersection followed by internal conversion to the ground state. A dedicated analysis of non-adiabatic effects, including the topology of the intersection seam, is therefore essential for a quantitative assessment of the relaxation pathways. Addressing these aspects in future work will be important for further optimisation of the system.

In addition, substitution may increase the relevance of competing deactivation channels, which reinforces the need for an explicit excited-state dynamics perspective beyond minimum-energy pathways alone. More broadly, it will be of interest to test whether the favourable combination of MOST-relevant criteria and \ce{BN}-enabled relaxation channels persists in larger \ce{BN}-bridged hydrocarbons, for example in anthracene and tetracene derivatives, and to evaluate how these trends evolve with increasing conjugation length.

\section*{Author Contributions}
M. B.: writing – original draft, writing – review \& editing, conceptualisation, investigation, methodology, validation, visualisation; M. I. S. R.: writing – review \& editing, conceptualisation, project administration, supervision, funding acquisition.

\section*{Conflicts of interest}
The authors declare no conflicts of interest.

\section*{Data availability}
The data and computational input files underlying this study are openly available, including NEB isomerisation pathways, XMS-CASPT2 single-point calculations, NICS input files, molecular structures, optimised transition states and the corresponding data for the substituted systems. These resources can be found in the GitHub repository located at \href{https://github.com/roehr-lab/BN_Dewar}{\url{https://github.com/roehr-lab/BN_Dewar}}. Users can freely access and download the materials under the terms of the repository’s license to facilitate further research and verification of the results presented herein.
\section*{Acknowledgements}
The authors gratefully acknowledge Michał Andrzej Kochman for a valuable discussion during a conference. M. B. and M. I. S. R. are thankful for financial support by the Deutsche Forschungsgemeinschaft (DFG, German Research Foundation) – Project-ID 551403841 – SFB 1762 “Boron as Property-Determining Element”.

%%%REFERENCES%%%
\bibliography{BN_DEWAR} %You need to replace "rsc" on this line with the name of your .bib file

@article{stojanovicMonoBNsubstitutedAnalogues2018,
	title = {Mono {BN}-substituted analogues of naphthalene: a theoretical analysis of the effect of {BN} position on stability, aromaticity and frontier orbital energies},
	volume = {42},
	issn = {1369-9261},
	shorttitle = {Mono {BN}-substituted analogues of naphthalene},
	url = {https://pubs.rsc.org/en/content/articlelanding/2018/nj/c8nj01529e},
	doi = {10.1039/C8NJ01529E},
	language = {en},
	number = {15},
	urldate = {2026-05-08},
	journal = {New Journal of Chemistry},
	publisher = {The Royal Society of Chemistry},
	author = {Stojanović, Milovan and Baranac-Stojanović, Marija},
	month = jul,
	year = {2018},
	pages = {12968--12976},
}

@article{appiariusBNSubstitutionDithienylpyrenesPrevents2022,
	title = {{BN}-{Substitution} in {Dithienylpyrenes} {Prevents} {Excimer} {Formation} in {Solution} and in the {Solid} {State}},
	volume = {126},
	issn = {1932-7447},
	url = {https://doi.org/10.1021/acs.jpcc.1c08812},
	doi = {10.1021/acs.jpcc.1c08812},
	number = {9},
	urldate = {2026-05-08},
	journal = {The Journal of Physical Chemistry C},
	publisher = {American Chemical Society},
	author = {Appiarius, Yannik and Gliese, Philipp J. and Segler, Stephan A. W. and Rusch, Pascal and Zhang, Jiangbin and Gates, Paul J. and Pal, Rumpa and Malaspina, Lorraine A. and Sugimoto, Kunihisa and Neudecker, Tim and Bigall, Nadja C. and Grabowsky, Simon and Bakulin, Artem A. and Staubitz, Anne},
	month = mar,
	year = {2022},
	pages = {4563--4576},
}

@article{schottlerLongTermEnergyStorage2022,
	title = {Long-{Term} {Energy} {Storage} {Systems} {Based} on the {Dihydroazulene}/{Vinylheptafulvene} {Photo}-/{Thermoswitch}},
	volume = {6},
	copyright = {© 2022 The Authors. ChemPhotoChem published by Wiley-VCH GmbH},
	issn = {2367-0932},
	url = {https://onlinelibrary.wiley.com/doi/abs/10.1002/cptc.202200037},
	doi = {10.1002/cptc.202200037},
	language = {en},
	number = {8},
	urldate = {2026-05-08},
	journal = {ChemPhotoChem},
	author = {Schøttler, Christina and Vegge, Siri Krogh and Cacciarini, Martina and Nielsen, Mogens Brøndsted},
	year = {2022},
	note = {\_eprint: https://chemistry-europe.onlinelibrary.wiley.com/doi/pdf/10.1002/cptc.202200037},
	pages = {e202200037},
}

@article{lindbaekbromanDihydroazuleneControllingPhotochromism2014,
	title = {Dihydroazulene: from controlling photochromism to molecular electronics devices},
	volume = {16},
	shorttitle = {Dihydroazulene},
	url = {https://pubs.rsc.org/en/content/articlelanding/2014/cp/c4cp02442g},
	doi = {10.1039/C4CP02442G},
	language = {en},
	number = {39},
	urldate = {2026-05-08},
	journal = {Physical Chemistry Chemical Physics},
	publisher = {Royal Society of Chemistry},
	author = {Lindbæk Broman, Søren and Brøndsted Nielsen, Mogens},
	year = {2014},
	pages = {21172--21182},
}

@article{neeseSoftwareUpdateORCA2025,
	title = {Software {Update}: {The} {ORCA} {Program} {System}—{Version} 6.0},
	volume = {15},
	copyright = {© 2025 The Author(s). WIREs Computational Molecular Science published by Wiley Periodicals LLC.},
	issn = {1759-0884},
	shorttitle = {Software {Update}},
	url = {https://onlinelibrary.wiley.com/doi/abs/10.1002/wcms.70019},
	doi = {10.1002/wcms.70019},
	language = {en},
	number = {2},
	urldate = {2025-07-03},
	journal = {WIREs Computational Molecular Science},
	author = {Neese, Frank},
	year = {2025},
	pages = {e70019},
}

@article{ozakiPositionalIsomerization12Azaborine2026,
	title = {Positional {Isomerization} of 1,2-{Azaborine} through {BN}-{Benzvalene}},
	volume = {148},
	issn = {0002-7863},
	url = {https://doi.org/10.1021/jacs.5c21413},
	doi = {10.1021/jacs.5c21413},
	urldate = {2026-01-22},
	journal = {Journal of the American Chemical Society},
	publisher = {American Chemical Society},
	author = {Ozaki, Tomoya and Diamandis, Skylar and Rybansky, Nina and Yang, Xinyu and Bai, Fan and Li, Bo and Huang, Jier and Liu, Shih-Yuan},
	month = jan,
	year = {2026},
	pages = {3820--3829},
}

@article{suMechanisticInvestigationsPhotoisomerization2013,
	title = {Mechanistic {Investigations} on the {Photoisomerization} {Reactions} of 1,2-{Dihydro}-1,2-{Azaborine}},
	volume = {19},
	copyright = {Copyright © 2013 WILEY-VCH Verlag GmbH \& Co. KGaA, Weinheim},
	issn = {1521-3765},
	url = {https://onlinelibrary.wiley.com/doi/abs/10.1002/chem.201204537},
	doi = {10.1002/chem.201204537},
	language = {en},
	number = {29},
	urldate = {2026-01-21},
	journal = {Chemistry – A European Journal},
	author = {Su, Ming-Der},
	year = {2013},
	note = {\_eprint: https://chemistry-europe.onlinelibrary.wiley.com/doi/pdf/10.1002/chem.201204537},
	pages = {9663--9667},
}

@article{kochmanSimulatingEnergyCapture2025,
	title = {Simulating the {Energy} {Capture} {Process} in {Push}–{Pull} {Norbornadiene}-{Quadricyclane} {Photoswitches}},
	volume = {16},
	url = {https://doi.org/10.1021/acs.jpclett.5c00634},
	doi = {10.1021/acs.jpclett.5c00634},
	number = {17},
	urldate = {2026-05-05},
	journal = {The Journal of Physical Chemistry Letters},
	publisher = {American Chemical Society},
	author = {Kochman, Michał Andrzej and Durbeej, Bo},
	month = may,
	year = {2025},
	pages = {4315--4325},
}

@article{rulliCatellaniInspiredBNAromaticExpansion2026,
	title = {Catellani-{Inspired} {BN}-{Aromatic} {Expansion}: {A} {Versatile} {Tool} toward π-{Extended} 1,2-{Azaborines} with {Tunable} {Photosensitizing} {Properties}},
	volume = {148},
	issn = {0002-7863},
	shorttitle = {Catellani-{Inspired} {BN}-{Aromatic} {Expansion}},
	url = {https://doi.org/10.1021/jacs.5c19389},
	doi = {10.1021/jacs.5c19389},
	number = {3},
	urldate = {2026-04-13},
	journal = {Journal of the American Chemical Society},
	publisher = {American Chemical Society},
	author = {Rulli, Federica and Ordeix, Sergi and Bresolí-Obach, Roger and Nonell, Santi and Saurí, Josep and Ribas-Font, Cristina and Shafir, Alexandr and Puig de la Bellacasa, Raimon and Cuenca, Ana B.},
	month = jan,
	year = {2026},
	pages = {3614--3625},
}

@article{nguyenMolecularSolarThermal2026,
	title = {Molecular solar thermal energy storage in {Dewar} pyrimidone beyond 1.6 {MJ}/kg},
	volume = {0},
	url = {https://www.science.org/doi/10.1126/science.aec6413},
	doi = {10.1126/science.aec6413},
	number = {0},
	urldate = {2026-03-23},
	journal = {Science},
	publisher = {American Association for the Advancement of Science},
	author = {Nguyen, Han P. Q. and Maertens, Alexander J. and Baker, Benjamin A. and Wu, Nathan M.-W. and Ye, Zihao and Zhou, Qingyang and Qiu, Qianfeng and Kaur, Navneet and Berkinsky, David B. and Shulenberger, Katherine E. and Houk, K. N. and Han, Grace G. D.},
	month = feb,
	year = {2026},
	pages = {eaec6413},
}

@article{schatzMolecularSolarThermal2026,
	title = {Molecular {Solar} {Thermal} ({MOST}) {Energy} {Storage}—{Definitions} and {Requirements} {Revisited}},
	volume = {65},
	copyright = {© 2025 The Author(s). Angewandte Chemie International Edition published by Wiley-VCH GmbH},
	issn = {1521-3773},
	url = {https://onlinelibrary.wiley.com/doi/abs/10.1002/anie.202520673},
	doi = {10.1002/anie.202520673},
	language = {en},
	number = {1},
	urldate = {2026-03-23},
	journal = {Angewandte Chemie International Edition},
	author = {Schatz, Dominic and Wegner, Hermann A.},
	year = {2026},
	note = {\_eprint: https://onlinelibrary.wiley.com/doi/pdf/10.1002/anie.202520673},
	pages = {e20673},
}

@article{sunApplicationsPhotoswitchesStorage2019,
	title = {Applications of {Photoswitches} in the {Storage} of {Solar} {Energy}},
	volume = {3},
	copyright = {© 2019 Wiley-VCH Verlag GmbH \& Co. KGaA, Weinheim},
	issn = {2367-0932},
	url = {https://onlinelibrary.wiley.com/doi/abs/10.1002/cptc.201900030},
	doi = {10.1002/cptc.201900030},
	language = {en},
	number = {6},
	urldate = {2026-01-21},
	journal = {ChemPhotoChem},
	author = {Sun, Cai-Li and Wang, Chenxu and Boulatov, Roman},
	year = {2019},
	note = {\_eprint: https://chemistry-europe.onlinelibrary.wiley.com/doi/pdf/10.1002/cptc.201900030},
	pages = {268--283},
}

@article{jeongUltrafastPhotoisomerizationMechanism2023,
	title = {Ultrafast photoisomerization mechanism of azaborine revealed by nonadiabatic molecular dynamics simulations},
	volume = {25},
	issn = {1463-9084},
	url = {https://pubs.rsc.org/en/content/articlelanding/2023/cp/d3cp01169k},
	doi = {10.1039/D3CP01169K},
	language = {en},
	number = {26},
	urldate = {2026-01-22},
	journal = {Physical Chemistry Chemical Physics},
	publisher = {The Royal Society of Chemistry},
	author = {Jeong, Sangmin and Park, Eunji and Kim, Joonghan and Kim, Kyung Hwan},
	month = jul,
	year = {2023},
	pages = {17230--17237},
}

@article{ozakiBoronNitrogenContainingBenzeneValence2024,
	title = {Boron-{Nitrogen}-{Containing} {Benzene} {Valence} {Isomers}},
	volume = {30},
	copyright = {© 2024 The Author(s). Chemistry - A European Journal published by Wiley-VCH GmbH},
	issn = {1521-3765},
	url = {https://onlinelibrary.wiley.com/doi/abs/10.1002/chem.202402544},
	doi = {10.1002/chem.202402544},
	language = {en},
	number = {55},
	urldate = {2026-01-20},
	journal = {Chemistry – A European Journal},
	author = {Ozaki, Tomoya and Liu, Shih-Yuan},
	year = {2024},
	note = {\_eprint: https://chemistry-europe.onlinelibrary.wiley.com/doi/pdf/10.1002/chem.202402544},
	pages = {e202402544},
}

@article{bettingerRingOpening2aza3borabicyclo220hex5ene2013,
	title = {Ring opening of 2-aza-3-borabicyclo[2.2.0]hex-5-ene, the {Dewar} form of 1,2-dihydro-1,2-azaborine: stepwise versus concerted mechanisms},
	volume = {9},
	copyright = {© 2013 Bettinger and Hauler; licensee Beilstein-Institut.},
	issn = {1860-5397},
	shorttitle = {Ring opening of 2-aza-3-borabicyclo[2.2.0]hex-5-ene, the {Dewar} form of 1,2-dihydro-1,2-azaborine},
	url = {https://www.beilstein-journals.org/bjoc/articles/9/86},
	doi = {10.3762/bjoc.9.86},
	language = {en},
	number = {1},
	urldate = {2026-01-21},
	journal = {Beilstein Journal of Organic Chemistry},
	publisher = {Beilstein-Institut},
	author = {Bettinger, Holger F. and Hauler, Otto},
	month = apr,
	year = {2013},
	pages = {761--766},
}

@article{campbellRecentAdvancesAzaborine2012,
	title = {Recent {Advances} in {Azaborine} {Chemistry}},
	volume = {51},
	copyright = {Copyright © 2012 WILEY-VCH Verlag GmbH \& Co. KGaA, Weinheim},
	issn = {1521-3773},
	url = {https://onlinelibrary.wiley.com/doi/abs/10.1002/anie.201200063},
	doi = {10.1002/anie.201200063},
	language = {en},
	number = {25},
	urldate = {2025-12-15},
	journal = {Angewandte Chemie International Edition},
	author = {Campbell, Patrick G. and Marwitz, Adam J. V. and Liu, Shih-Yuan},
	year = {2012},
	note = {\_eprint: https://onlinelibrary.wiley.com/doi/pdf/10.1002/anie.201200063},
	pages = {6074--6092},
}

@article{broughPhotoisomerization12Dihydro12AzaborineMatrix2012,
	title = {Photoisomerization of 1,2-{Dihydro}-1,2-{Azaborine}: {A} {Matrix} {Isolation} {Study}},
	volume = {51},
	copyright = {Copyright © 2012 WILEY-VCH Verlag GmbH \& Co. KGaA, Weinheim},
	issn = {1521-3773},
	shorttitle = {Photoisomerization of 1,2-{Dihydro}-1,2-{Azaborine}},
	url = {https://onlinelibrary.wiley.com/doi/abs/10.1002/anie.201203546},
	doi = {10.1002/anie.201203546},
	number = {43},
	urldate = {2025-12-15},
	journal = {Angewandte Chemie International Edition},
	author = {Brough, Sarah A. and Lamm, Ashley N. and Liu, Shih-Yuan and Bettinger, Holger F.},
	year = {2012},
	note = {\_eprint: https://onlinelibrary.wiley.com/doi/pdf/10.1002/anie.201203546},
	pages = {10880--10883},
}

@article{latajkaCNDO2Molecular1981,
	title = {{CNDO}/2 molecular orbital calculations of the hemi-{Dewar} and valene structures of naphthalene},
	volume = {86},
	issn = {0166-1280},
	url = {https://www.sciencedirect.com/science/article/pii/0166128081850737},
	doi = {10.1016/0166-1280(81)85073-7},
	number = {1},
	urldate = {2026-02-24},
	journal = {Journal of Molecular Structure: THEOCHEM},
	author = {Latajka, Zdzisław and Ratajczak, Henryk and Orville-Thomas, W. J. and Ratajczak, Emil},
	month = dec,
	year = {1981},
	pages = {91--95},
}

@article{mcconnellLatestageFunctionalizationBNheterocycles2019,
	title = {Late-stage functionalization of {BN}-heterocycles},
	volume = {48},
	issn = {1460-4744},
	url = {https://pubs.rsc.org/en/content/articlelanding/2019/cs/c9cs00218a},
	doi = {10.1039/C9CS00218A},
	language = {en},
	number = {13},
	urldate = {2025-09-30},
	journal = {Chemical Society Reviews},
	publisher = {The Royal Society of Chemistry},
	author = {McConnell, Cameron R. and Liu, Shih-Yuan},
	month = jul,
	year = {2019},
	pages = {3436--3453},
}

@article{giustraStateArtAzaborine2018,
	title = {The {State} of the {Art} in {Azaborine} {Chemistry}: {New} {Synthetic} {Methods} and {Applications}},
	volume = {140},
	issn = {0002-7863},
	shorttitle = {The {State} of the {Art} in {Azaborine} {Chemistry}},
	url = {https://doi.org/10.1021/jacs.7b09446},
	doi = {10.1021/jacs.7b09446},
	number = {4},
	urldate = {2025-09-30},
	journal = {Journal of the American Chemical Society},
	publisher = {American Chemical Society},
	author = {Giustra, Zachary X. and Liu, Shih-Yuan},
	month = jan,
	year = {2018},
	pages = {1184--1194},
}

@article{ishibashiBNTetraceneExtending2017,
	title = {{BN} {Tetracene}: {Extending} the {Reach} of {BN}/{CC} {Isosterism} in {Acenes}},
	volume = {36},
	issn = {0276-7333},
	shorttitle = {{BN} {Tetracene}},
	url = {https://doi.org/10.1021/acs.organomet.7b00296},
	doi = {10.1021/acs.organomet.7b00296},
	number = {14},
	urldate = {2026-03-10},
	journal = {Organometallics},
	publisher = {American Chemical Society},
	author = {Ishibashi, Jacob S. A. and Dargelos, Alain and Darrigan, Clovis and Chrostowska, Anna and Liu, Shih-Yuan},
	month = jul,
	year = {2017},
	pages = {2494--2497},
}

@article{rulliPropenolysisEnyneMetathesis2024,
	title = {From propenolysis to enyne metathesis: tools for expedited assembly of 4a,8a-azaboranaphthalene and extended polycycles with embedded {BN}},
	volume = {15},
	issn = {2041-6539},
	shorttitle = {From propenolysis to enyne metathesis},
	url = {https://pubs.rsc.org/en/content/articlelanding/2024/sc/d3sc06676b},
	doi = {10.1039/D3SC06676B},
	language = {en},
	number = {15},
	urldate = {2026-03-10},
	journal = {Chemical Science},
	publisher = {The Royal Society of Chemistry},
	author = {Rulli, Federica and Sanz-Liarte, Guillem and Roca, Pol and Martínez, Nina and Medina, Víctor and Bellacasa, Raimon Puig de la and Shafir, Alexandr and Cuenca, Ana B.},
	month = apr,
	year = {2024},
	pages = {5674--5680},
}

@article{wangPyAromaIntuitiveGraphical2024,
	title = {py.{Aroma}: {An} {Intuitive} {Graphical} {User} {Interface} for {Diverse} {Aromaticity} {Analyses}},
	volume = {6},
	copyright = {http://creativecommons.org/licenses/by/3.0/},
	issn = {2624-8549},
	shorttitle = {py.{Aroma}},
	url = {https://www.mdpi.com/2624-8549/6/6/103},
	doi = {10.3390/chemistry6060103},
	language = {en},
	number = {6},
	urldate = {2026-02-24},
	journal = {Chemistry},
	publisher = {Multidisciplinary Digital Publishing Institute},
	author = {Wang, Zhe},
	month = dec,
	year = {2024},
	pages = {1692--1703},
}

@article{guestenDeactivationFluorescentState1980,
	title = {Deactivation of the fluorescent state of 9-tert-butylanthracene. 9-tert-{Butyl}-9,10({Dewar} anthracene)},
	volume = {102},
	issn = {0002-7863, 1520-5126},
	url = {https://pubs.acs.org/doi/abs/10.1021/ja00547a022},
	doi = {10.1021/ja00547a022},
	language = {en},
	number = {27},
	urldate = {2026-02-24},
	journal = {Journal of the American Chemical Society},
	author = {Guesten, H. and Mintas, M. and Klasinc, L.},
	month = dec,
	year = {1980},
	pages = {7936--7937},
}

@article{qiuVisibleLightActivated2023,
	title = {Visible light activated energy storage in solid-state {Azo}-{BF2} switches},
	volume = {14},
	issn = {2041-6539},
	url = {https://pubs.rsc.org/en/content/articlelanding/2023/sc/d3sc03465h},
	doi = {10.1039/D3SC03465H},
	language = {en},
	number = {41},
	urldate = {2026-02-24},
	journal = {Chemical Science},
	publisher = {The Royal Society of Chemistry},
	author = {Qiu, Qianfeng and Qi, Qingkai and Usuba, Junichi and Lee, Karina and Aprahamian, Ivan and Han, Grace G. D.},
	month = oct,
	year = {2023},
	pages = {11359--11364},
}

@article{jornerUnravelingFactorsLeading2017,
	title = {Unraveling factors leading to efficient norbornadiene–quadricyclane molecular solar-thermal energy storage systems},
	volume = {5},
	issn = {2050-7496},
	url = {https://pubs.rsc.org/en/content/articlelanding/2017/ta/c7ta04259k},
	doi = {10.1039/C7TA04259K},
	language = {en},
	number = {24},
	urldate = {2026-02-24},
	journal = {Journal of Materials Chemistry A},
	publisher = {The Royal Society of Chemistry},
	author = {Jorner, Kjell and Dreos, Ambra and Emanuelsson, Rikard and Bakouri, Ouissam El and Galván, Ignacio Fdez and Börjesson, Karl and Feixas, Ferran and Lindh, Roland and Zietz, Burkhard and Moth-Poulsen, Kasper and Ottosson, Henrik},
	month = jun,
	year = {2017},
	pages = {12369--12378},
}

@article{gimenez-gomezStateoftheartChallengesMolecular2024,
	title = {State-of-the-art and challenges towards a {Molecular} {Solar} {Thermal} ({MOST}) energy storage device},
	volume = {9},
	issn = {2058-9883},
	url = {https://pubs.rsc.org/en/content/articlelanding/2024/re/d4re00131a},
	doi = {10.1039/D4RE00131A},
	language = {en},
	number = {7},
	urldate = {2026-02-24},
	journal = {Reaction Chemistry \& Engineering},
	publisher = {The Royal Society of Chemistry},
	author = {Giménez-Gómez, Alberto and Magson, Lucien and Merino-Robledillo, Cecilia and Hernáez-Troya, Sara and Sanosa, Nil and Sampedro, Diego and Funes-Ardoiz, Ignacio},
	month = jun,
	year = {2024},
	pages = {1629--1640},
}

@article{mikiSynthesisPhotoreaction1234Tetratbutylnaphthalene1992,
	title = {Synthesis and {Photoreaction} of 1,2,3,4-{Tetra}-t-butylnaphthalene: {A} {Highly} {Crowded} {Naphthalene} {Derivative} and {Its} {Valenceisomers}},
	volume = {33},
	issn = {0040-4039},
	shorttitle = {Synthesis and {Photoreaction} of 1,2,3,4-{Tetra}-t-butylnaphthalene},
	url = {https://www.sciencedirect.com/science/article/pii/S0040403900916897},
	doi = {10.1016/S0040-4039(00)91689-7},
	number = {12},
	urldate = {2026-02-24},
	journal = {Tetrahedron Letters},
	author = {Miki, Sadao and Ema, Tadashi and Shimizu, Rie and Nakatsuji, Hiroshi and Yoshida, Zen-ichi},
	month = mar,
	year = {1992},
	pages = {1619--1620},
}

@article{chakrabortyCurvedAnthracenesVisiblelight2025,
	title = {Curved anthracenes for visible-light photon energy storage via {Dewar} isomerization},
	volume = {11},
	issn = {2451-9294},
	url = {https://www.sciencedirect.com/science/article/pii/S2451929425002517},
	doi = {10.1016/j.chempr.2025.102660},
	number = {12},
	urldate = {2026-02-24},
	journal = {Chem},
	author = {Chakraborty, Subhayan and Choudhuri, Writam S. R. and Usuba, Junichi and Qiu, Qianfeng and Raju, Cijil and Han, Grace G. D.},
	month = dec,
	year = {2025},
	pages = {102660},
}

@article{grayDiarylsubstitutedNorbornadienesRedshifted2014,
	title = {Diaryl-substituted norbornadienes with red-shifted absorption for molecular solar thermal energy storage},
	volume = {50},
	issn = {1364-548X},
	url = {https://pubs.rsc.org/en/content/articlelanding/2014/cc/c3cc47517d},
	doi = {10.1039/C3CC47517D},
	language = {en},
	number = {40},
	urldate = {2026-02-24},
	journal = {Chemical Communications},
	publisher = {The Royal Society of Chemistry},
	author = {Gray, Victor and Lennartson, Anders and Ratanalert, Phasin and Börjesson, Karl and Moth-Poulsen, Kasper},
	month = apr,
	year = {2014},
	pages = {5330--5332},
}

@article{kuismaComparativeAbInitioStudy2016,
	title = {Comparative {Ab}-{Initio} {Study} of {Substituted} {Norbornadiene}-{Quadricyclane} {Compounds} for {Solar} {Thermal} {Storage}},
	volume = {120},
	issn = {1932-7447},
	url = {https://doi.org/10.1021/acs.jpcc.5b11489},
	doi = {10.1021/acs.jpcc.5b11489},
	number = {7},
	urldate = {2026-02-24},
	journal = {The Journal of Physical Chemistry C},
	publisher = {American Chemical Society},
	author = {Kuisma, Mikael J. and Lundin, Angelica M. and Moth-Poulsen, Kasper and Hyldgaard, Per and Erhart, Paul},
	month = feb,
	year = {2016},
	pages = {3635--3645},
}

@article{franzElectrochemicallyTriggeredEnergy2022,
	title = {Electrochemically {Triggered} {Energy} {Release} from an {Azothiophene}-{Based} {Molecular} {Solar} {Thermal} {System}},
	volume = {15},
	copyright = {© 2022 The Authors. ChemSusChem published by Wiley-VCH GmbH},
	issn = {1864-564X},
	url = {https://onlinelibrary.wiley.com/doi/abs/10.1002/cssc.202200958},
	doi = {10.1002/cssc.202200958},
	language = {en},
	number = {18},
	urldate = {2026-02-24},
	journal = {ChemSusChem},
	author = {Franz, Evanie and Kunz, Anne and Oberhof, Nils and Heindl, Andreas H. and Bertram, Manon and Fusek, Lukas and Taccardi, Nicola and Wasserscheid, Peter and Dreuw, Andreas and Wegner, Hermann A. and Brummel, Olaf and Libuda, Jörg},
	year = {2022},
	note = {\_eprint: https://chemistry-europe.onlinelibrary.wiley.com/doi/pdf/10.1002/cssc.202200958},
	pages = {e202200958},
}

@article{chakrabortySelfactivatedEnergyRelease2024,
	title = {Self-activated energy release cascade from anthracene-based solid-state molecular solar thermal energy storage systems},
	volume = {10},
	issn = {2451-9294},
	url = {https://www.sciencedirect.com/science/article/pii/S2451929424003127},
	doi = {10.1016/j.chempr.2024.06.033},
	number = {11},
	urldate = {2026-02-24},
	journal = {Chem},
	author = {Chakraborty, Subhayan and Nguyen, Han P. Q. and Usuba, Junichi and Choi, Ji Yong and Sun, Zhenhuan and Raju, Cijil and Sigelmann, Gustavo and Qiu, Qianfeng and Cho, Sungwon and Tenney, Stephanie M. and Shulenberger, Katherine E. and Schmidt-Rohr, Klaus and Park, Jihye and Han, Grace G. D.},
	month = nov,
	year = {2024},
	pages = {3309--3322},
}

@article{borjessonEfficiencyLimitMolecular2013,
	title = {Efficiency {Limit} of {Molecular} {Solar} {Thermal} {Energy} {Collecting} {Devices}},
	volume = {1},
	url = {https://doi.org/10.1021/sc300107z},
	doi = {10.1021/sc300107z},
	number = {6},
	urldate = {2026-02-24},
	journal = {ACS Sustainable Chemistry \& Engineering},
	publisher = {American Chemical Society},
	author = {Börjesson, Karl and Lennartson, Anders and Moth-Poulsen, Kasper},
	month = jun,
	year = {2013},
	pages = {585--590},
}

@article{marwitzHybridOrganicInorganic2009,
	title = {A {Hybrid} {Organic}/{Inorganic} {Benzene}},
	volume = {48},
	copyright = {Copyright © 2009 WILEY-VCH Verlag GmbH \& Co. KGaA, Weinheim},
	issn = {1521-3773},
	url = {https://onlinelibrary.wiley.com/doi/abs/10.1002/anie.200805554},
	doi = {10.1002/anie.200805554},
	number = {5},
	urldate = {2026-02-24},
	journal = {Angewandte Chemie International Edition},
	author = {Marwitz, Adam J. V. and Matus, Myrna H. and Zakharov, Lev N. and Dixon, David A. and Liu, Shih-Yuan},
	year = {2009},
	note = {\_eprint: https://onlinelibrary.wiley.com/doi/pdf/10.1002/anie.200805554},
	pages = {973--977},
}

@article{asgeirssonNudgedElasticBand2021,
	title = {Nudged {Elastic} {Band} {Method} for {Molecular} {Reactions} {Using} {Energy}-{Weighted} {Springs} {Combined} with {Eigenvector} {Following}},
	volume = {17},
	issn = {1549-9618},
	url = {https://doi.org/10.1021/acs.jctc.1c00462},
	doi = {10.1021/acs.jctc.1c00462},
	number = {8},
	urldate = {2026-02-23},
	journal = {Journal of Chemical Theory and Computation},
	publisher = {American Chemical Society},
	author = {Ásgeirsson, Vilhjálmur and Birgisson, Benedikt Orri and Bjornsson, Ragnar and Becker, Ute and Neese, Frank and Riplinger, Christoph and Jónsson, Hannes},
	month = aug,
	year = {2021},
	pages = {4929--4945},
}

@article{orrego-hernandezEngineeringNorbornadieneQuadricyclane2020,
	title = {Engineering of {Norbornadiene}/{Quadricyclane} {Photoswitches} for {Molecular} {Solar} {Thermal} {Energy} {Storage} {Applications}},
	volume = {53},
	issn = {0001-4842},
	url = {https://doi.org/10.1021/acs.accounts.0c00235},
	doi = {10.1021/acs.accounts.0c00235},
	number = {8},
	urldate = {2026-02-23},
	journal = {Accounts of Chemical Research},
	publisher = {American Chemical Society},
	author = {Orrego-Hernández, Jessica and Dreos, Ambra and Moth-Poulsen, Kasper},
	month = aug,
	year = {2020},
	pages = {1478--1487},
}

@article{hemauerNorbornadieneQuadricyclanePair2024,
	title = {The {Norbornadiene}/{Quadricyclane} {Pair} as {Molecular} {Solar} {Thermal} {Energy} {Storage} {System}: {Surface} {Science} {Investigations}},
	volume = {25},
	copyright = {© 2024 The Authors. ChemPhysChem published by Wiley-VCH GmbH},
	issn = {1439-7641},
	shorttitle = {The {Norbornadiene}/{Quadricyclane} {Pair} as {Molecular} {Solar} {Thermal} {Energy} {Storage} {System}},
	url = {https://onlinelibrary.wiley.com/doi/abs/10.1002/cphc.202300806},
	doi = {10.1002/cphc.202300806},
	language = {en},
	number = {9},
	urldate = {2026-02-23},
	journal = {ChemPhysChem},
	author = {Hemauer, Felix and Steinrück, Hans-Peter and Papp, Christian},
	year = {2024},
	note = {\_eprint: https://chemistry-europe.onlinelibrary.wiley.com/doi/pdf/10.1002/cphc.202300806},
	pages = {e202300806},
}

@article{gimenez-gomezPhotochemicalOverviewMolecular2022,
	title = {A {Photochemical} {Overview} of {Molecular} {Solar} {Thermal} {Energy} {Storage}},
	volume = {2},
	copyright = {http://creativecommons.org/licenses/by/3.0/},
	issn = {2673-7256},
	url = {https://www.mdpi.com/2673-7256/2/3/45},
	doi = {10.3390/photochem2030045},
	language = {en},
	number = {3},
	urldate = {2026-02-23},
	journal = {Photochem},
	publisher = {Multidisciplinary Digital Publishing Institute},
	author = {Gimenez-Gomez, Alberto and Magson, Lucien and Peñin, Beatriz and Sanosa, Nil and Soilán, Jacobo and Losantos, Raúl and Sampedro, Diego},
	month = sep,
	year = {2022},
	pages = {694--716},
}

@article{m.arpaPhotochemicalFormationElusive2024,
	title = {Photochemical formation of the elusive {Dewar} isomers of aromatic systems: why are substituted azaborines different?},
	volume = {26},
	shorttitle = {Photochemical formation of the elusive {Dewar} isomers of aromatic systems},
	url = {https://pubs.rsc.org/en/content/articlelanding/2024/cp/d4cp00777h},
	doi = {10.1039/D4CP00777H},
	language = {en},
	number = {15},
	urldate = {2026-01-20},
	journal = {Physical Chemistry Chemical Physics},
	publisher = {Royal Society of Chemistry},
	author = {M. Arpa, Enrique and Stafström, Sven and Durbeej, Bo},
	year = {2024},
	pages = {11295--11305},
}

@article{edelDewarIsomer12Dihydro12azaborinines2018,
	title = {The {Dewar} {Isomer} of 1,2-{Dihydro}-1,2-azaborinines: {Isolation}, {Fragmentation}, and {Energy} {Storage}},
	volume = {57},
	copyright = {© 2018 Wiley-VCH Verlag GmbH \& Co. KGaA, Weinheim},
	issn = {1521-3773},
	shorttitle = {The {Dewar} {Isomer} of 1,2-{Dihydro}-1,2-azaborinines},
	url = {https://onlinelibrary.wiley.com/doi/abs/10.1002/anie.201712683},
	doi = {10.1002/anie.201712683},
	language = {en},
	number = {19},
	urldate = {2026-01-21},
	journal = {Angewandte Chemie International Edition},
	author = {Edel, Klara and Yang, Xinyu and Ishibashi, Jacob S. A. and Lamm, Ashley N. and Maichle-Mössmer, Cäcilia and Giustra, Zachary X. and Liu, Shih-Yuan and Bettinger, Holger F.},
	year = {2018},
	note = {\_eprint: https://onlinelibrary.wiley.com/doi/pdf/10.1002/anie.201712683},
	pages = {5296--5300},
}

@article{guerraBNDewarBenzeneBNbenzvalene2025,
	title = {From {BN}-{Dewar} benzene to {BN}-benzvalene: a computational exploration of photoisomerization mechanisms},
	volume = {23},
	issn = {1477-0539},
	shorttitle = {From {BN}-{Dewar} benzene to {BN}-benzvalene},
	url = {https://pubs.rsc.org/en/content/articlelanding/2025/ob/d5ob01156f},
	doi = {10.1039/D5OB01156F},
	language = {en},
	number = {38},
	urldate = {2026-01-20},
	journal = {Organic \& Biomolecular Chemistry},
	publisher = {The Royal Society of Chemistry},
	author = {Guerra, Cristian J. and Rodríguez-Núñez, Yeray A. and Polo-Cuadrado, Efraín and Ayarde-Henríquez, Leandro and Ramírez, Diana B. and Ensuncho, Adolfo E.},
	month = oct,
	year = {2025},
	pages = {8769--8777},
}

@article{ozakiBNBenzvalene2024,
	title = {A {BN}-{Benzvalene}},
	volume = {146},
	issn = {0002-7863},
	url = {https://doi.org/10.1021/jacs.4c08088},
	doi = {10.1021/jacs.4c08088},
	number = {36},
	urldate = {2026-01-20},
	journal = {Journal of the American Chemical Society},
	publisher = {American Chemical Society},
	author = {Ozaki, Tomoya and Bentley, Sierra K. and Rybansky, Nina and Li, Bo and Liu, Shih-Yuan},
	month = sep,
	year = {2024},
	pages = {24748--24753},
}

@article{kimTheoreticalInvestigationReaction2015,
	title = {Theoretical {Investigation} of the {Reaction} {Mechanism} of the {Photoisomerization} of 1,2-{Dihydro}-1,2-azaborine},
	volume = {16},
	copyright = {© 2015 WILEY-VCH Verlag GmbH \& Co. KGaA, Weinheim},
	issn = {1439-7641},
	url = {https://onlinelibrary.wiley.com/doi/abs/10.1002/cphc.201500153},
	doi = {10.1002/cphc.201500153},
	language = {en},
	number = {8},
	urldate = {2026-01-20},
	journal = {ChemPhysChem},
	author = {Kim, Joonghan and Moon, Jiwon and Lim, Jeong Sik},
	year = {2015},
	note = {\_eprint: https://chemistry-europe.onlinelibrary.wiley.com/doi/pdf/10.1002/cphc.201500153},
	pages = {1670--1675},
}

@article{bieblHighEnergyDensity2025,
	title = {High energy density dihydroazaborinine dyads and triad for molecular solar thermal energy storage},
	volume = {16},
	issn = {2041-6539},
	url = {https://pubs.rsc.org/en/content/articlelanding/2025/sc/d5sc03159a},
	doi = {10.1039/D5SC03159A},
	language = {en},
	number = {33},
	urldate = {2026-01-20},
	journal = {Chemical Science},
	publisher = {The Royal Society of Chemistry},
	author = {Biebl, Sonja M. and Richter, Robert C. and Ströbele, Markus and Fleischer, Ivana and Bettinger, Holger F.},
	month = aug,
	year = {2025},
	pages = {15231--15238},
}

@article{j.mullerRationalDesignRedshifted2025,
	title = {Rational design of red-shifted 1,2-azaborinine-based molecular solar thermals},
	volume = {61},
	url = {https://pubs.rsc.org/en/content/articlelanding/2025/cc/d5cc01963j},
	doi = {10.1039/D5CC01963J},
	language = {en},
	number = {46},
	urldate = {2026-01-20},
	journal = {Chemical Communications},
	publisher = {Royal Society of Chemistry},
	author = {J. Müller, Adrian and Markhart, Johannes and F. Bettinger, Holger and Dreuw, Andreas},
	year = {2025},
	pages = {8351--8354},
}

@article{boelkeDesigningMolecularPhotoswitches2019,
	title = {Designing {Molecular} {Photoswitches} for {Soft} {Materials} {Applications}},
	volume = {7},
	copyright = {© 2019 WILEY-VCH Verlag GmbH \& Co. KGaA, Weinheim},
	issn = {2195-1071},
	url = {https://onlinelibrary.wiley.com/doi/abs/10.1002/adom.201900404},
	doi = {10.1002/adom.201900404},
	language = {en},
	number = {16},
	urldate = {2026-02-23},
	journal = {Advanced Optical Materials},
	author = {Boelke, Jan and Hecht, Stefan},
	year = {2019},
	note = {\_eprint: https://advanced.onlinelibrary.wiley.com/doi/pdf/10.1002/adom.201900404},
	pages = {1900404},
}

@article{kobauriRationalDesignPhotopharmacology2023,
	title = {Rational {Design} in {Photopharmacology} with {Molecular} {Photoswitches}},
	volume = {62},
	copyright = {© 2023 The Authors. Angewandte Chemie International Edition published by Wiley-VCH GmbH},
	issn = {1521-3773},
	url = {https://onlinelibrary.wiley.com/doi/abs/10.1002/anie.202300681},
	doi = {10.1002/anie.202300681},
	language = {en},
	number = {30},
	urldate = {2026-02-23},
	journal = {Angewandte Chemie International Edition},
	author = {Kobauri, Piermichele and Dekker, Frank J. and Szymanski, Wiktor and Feringa, Ben L.},
	year = {2023},
	note = {\_eprint: https://onlinelibrary.wiley.com/doi/pdf/10.1002/anie.202300681},
	pages = {e202300681},
}

@article{xuMolecularSolarThermal2025,
	title = {Molecular solar thermal energy storage devices: toward a more sustainable future},
	volume = {18},
	shorttitle = {Molecular solar thermal energy storage devices},
	url = {https://pubs.rsc.org/en/content/articlelanding/2025/ee/d5ee02556g},
	doi = {10.1039/D5EE02556G},
	language = {en},
	number = {20},
	urldate = {2026-02-23},
	journal = {Energy \& Environmental Science},
	publisher = {Royal Society of Chemistry},
	author = {Xu, Xingtang and Li, Chonghua and Li, Wang and Feng, Jie and Li, Wen-Ying},
	year = {2025},
	pages = {8990--9017},
}

@article{johnstoneMatrixcontrolledPhotochemistryBenzene1991,
	title = {Matrix-controlled photochemistry of benzene and pyridine},
	volume = {95},
	issn = {0022-3654, 1541-5740},
	url = {https://pubs.acs.org/doi/abs/10.1021/j100154a033},
	doi = {10.1021/j100154a033},
	language = {en},
	number = {1},
	urldate = {2026-02-23},
	journal = {The Journal of Physical Chemistry},
	author = {Johnstone, Duncan E. and Sodeau, John R.},
	month = jan,
	year = {1991},
	pages = {165--169},
}

@article{pieriNonadiabaticNanoreactorAutomated2021,
	title = {The non-adiabatic nanoreactor: towards the automated discovery of photochemistry},
	volume = {12},
	shorttitle = {The non-adiabatic nanoreactor},
	url = {https://pubs.rsc.org/en/content/articlelanding/2021/sc/d1sc00775k},
	doi = {10.1039/D1SC00775K},
	language = {en},
	number = {21},
	urldate = {2026-02-23},
	journal = {Chemical Science},
	publisher = {Royal Society of Chemistry},
	author = {Pieri, Elisa and Lahana, Dean and M. Chang, Alexander and R. Aldaz, Cody and C. Thompson, Keiran and J. Martínez, Todd},
	year = {2021},
	pages = {7294--7307},
}

@article{wardVacuumUltravioletPhotolysis1968,
	title = {Vacuum ultraviolet photolysis of benzene},
	volume = {90},
	issn = {0002-7863, 1520-5126},
	url = {https://pubs.acs.org/doi/abs/10.1021/ja01022a002},
	doi = {10.1021/ja01022a002},
	language = {en},
	number = {20},
	urldate = {2026-02-23},
	journal = {Journal of the American Chemical Society},
	author = {Ward, Harold R. and Wishnok, John S.},
	month = sep,
	year = {1968},
	pages = {5353--5357},
}

@article{richterFacileEnergyRelease2024,
	title = {Facile {Energy} {Release} from {Substituted} {Dewar} {Isomers} of 1,2-{Dihydro}-1,2-{Azaborinines} {Catalyzed} by {Coinage} {Metal} {Lewis} {Acids}},
	volume = {136},
	copyright = {© 2024 The Authors. Angewandte Chemie published by Wiley-VCH GmbH},
	issn = {1521-3757},
	url = {https://onlinelibrary.wiley.com/doi/abs/10.1002/ange.202405818},
	doi = {10.1002/ange.202405818},
	language = {en},
	number = {30},
	urldate = {2026-01-22},
	journal = {Angewandte Chemie},
	author = {Richter, Robert C. and Biebl, Sonja M. and Einholz, Ralf and Walz, Johannes and Maichle-Mössmer, Cäcilia and Ströbele, Markus and Bettinger, Holger F. and Fleischer, Ivana},
	year = {2024},
	note = {\_eprint: https://onlinelibrary.wiley.com/doi/pdf/10.1002/ange.202405818},
	pages = {e202405818},
}

@article{liuBNCCHow2008,
	title = {B{N} versus {C}{C}: {How} {Similar} {Are} {They}?},
	volume = {47},
	issn = {1521-3773},
	shorttitle = {B{N} versus {C}{C}},
	url = {https://onlinelibrary.wiley.com/doi/abs/10.1002/anie.200703535},
	doi = {10.1002/anie.200703535},
	number = {2},
	urldate = {2022-09-19},
	journal = {Angewandte Chemie International Edition},
	author = {Liu, Zhiqiang and Marder, Todd B.},
	year = {2008},
	note = {Number: 2
\_eprint: https://onlinelibrary.wiley.com/doi/pdf/10.1002/anie.200703535},
	pages = {242--244},
}

@article{kawaguchiMaterialsBasedGraphite1997,
	title = {B/{C}/{N} {Materials} {Based} on the {Graphite} {Network}},
	volume = {9},
	issn = {1521-4095},
	url = {https://onlinelibrary.wiley.com/doi/abs/10.1002/adma.19970090805},
	doi = {10.1002/adma.19970090805},
	number = {8},
	urldate = {2022-09-19},
	journal = {Advanced Materials},
	author = {Kawaguchi, Masayuki},
	year = {1997},
	note = {Number: 8
\_eprint: https://onlinelibrary.wiley.com/doi/pdf/10.1002/adma.19970090805},
	pages = {615--625},
}

@article{bosdetBNCCSubstitute2009,
	title = {B-{N} as a {C}-{C} substitute in aromatic systems},
	volume = {87},
	issn = {0008-4042},
	url = {https://cdnsciencepub.com/doi/abs/10.1139/v08-110},
	doi = {10.1139/v08-110},
	number = {1},
	urldate = {2022-09-19},
	journal = {Canadian Journal of Chemistry},
	publisher = {NRC Research Press},
	author = {Bosdet, Michael J.D and Piers, Warren E},
	month = jan,
	year = {2009},
	note = {Number: 1},
	pages = {8--29},
}

@article{snyderExcitedStateDeactivationPathways2017,
	title = {Excited-{State} {Deactivation} {Pathways} and the {Photocyclization} of {BN}-{Doped} {Polyaromatics}},
	volume = {121},
	issn = {1089-5639},
	url = {https://doi.org/10.1021/acs.jpca.7b04878},
	doi = {10.1021/acs.jpca.7b04878},
	number = {27},
	urldate = {2025-12-15},
	journal = {The Journal of Physical Chemistry A},
	publisher = {American Chemical Society},
	author = {Snyder, Joshua A. and Grüninger, Peter and Bettinger, Holger F. and Bragg, Arthur E.},
	month = jul,
	year = {2017},
	pages = {5136--5146},
}

@article{snyderBNDopingPhotochemistry2017,
	title = {{BN} {Doping} and the {Photochemistry} of {Polyaromatic} {Hydrocarbons}: {Photocyclization} of {Hexaphenyl} {Benzene} and {Hexaphenyl} {Borazine}},
	volume = {121},
	issn = {1089-5639},
	shorttitle = {{BN} {Doping} and the {Photochemistry} of {Polyaromatic} {Hydrocarbons}},
	url = {https://doi.org/10.1021/acs.jpca.7b08190},
	doi = {10.1021/acs.jpca.7b08190},
	number = {44},
	urldate = {2025-12-15},
	journal = {The Journal of Physical Chemistry A},
	publisher = {American Chemical Society},
	author = {Snyder, Joshua A. and Grüninger, Peter and Bettinger, Holger F. and Bragg, Arthur E.},
	month = nov,
	year = {2017},
	pages = {8359--8367},
}

@article{zengSeekingSmallMolecules2014,
	title = {Seeking {Small} {Molecules} for {Singlet} {Fission}: {A} {Heteroatom} {Substitution} {Strategy}},
	volume = {136},
	issn = {0002-7863},
	shorttitle = {Seeking {Small} {Molecules} for {Singlet} {Fission}},
	url = {https://doi.org/10.1021/ja505275m},
	doi = {10.1021/ja505275m},
	number = {36},
	urldate = {2025-03-06},
	journal = {Journal of the American Chemical Society},
	publisher = {American Chemical Society},
	author = {Zeng, Tao and Ananth, Nandini and Hoffmann, Roald},
	month = sep,
	year = {2014},
	pages = {12638--12647},
}

@article{sturmImpactIsoelectronicSubstitution2024,
	title = {Impact of isoelectronic substitution on the excited state processes in polycyclic aromatic hydrocarbons: a joint experimental and theoretical study of 4a,8a-azaboranaphthalene},
	volume = {26},
	issn = {1463-9084},
	shorttitle = {Impact of isoelectronic substitution on the excited state processes in polycyclic aromatic hydrocarbons},
	url = {https://pubs.rsc.org/en/content/articlelanding/2024/cp/d3cp05508f},
	doi = {10.1039/D3CP05508F},
	language = {en},
	number = {9},
	urldate = {2025-03-05},
	journal = {Physical Chemistry Chemical Physics},
	publisher = {The Royal Society of Chemistry},
	author = {Sturm, Floriane and Bühler, Michael and Stapper, Christoph and Schneider, Johannes S. and Helten, Holger and Fischer, Ingo and Röhr, Merle I. S.},
	month = feb,
	year = {2024},
	pages = {7363--7370},
}

@article{pinheiroSystematicAnalysisExcitonic2020,
	title = {A systematic analysis of excitonic properties to seek optimal singlet fission: the {BN}-substitution patterns in tetracene},
	volume = {8},
	issn = {2050-7534},
	shorttitle = {A systematic analysis of excitonic properties to seek optimal singlet fission},
	url = {https://pubs.rsc.org/en/content/articlelanding/2020/tc/c9tc06581d},
	doi = {10.1039/C9TC06581D},
	language = {en},
	number = {23},
	urldate = {2025-09-30},
	journal = {Journal of Materials Chemistry C},
	publisher = {The Royal Society of Chemistry},
	author = {Pinheiro, Max and Machado, Francisco B. C. and Plasser, Felix and Aquino, Adélia J. A. and Lischka, Hans},
	month = jun,
	year = {2020},
	pages = {7793--7804},
}

@article{morganEfficientSyntheticMethods2016,
	title = {Efficient synthetic methods for the installation of boron–nitrogen bonds in conjugated organic molecules},
	volume = {45},
	issn = {1477-9234},
	url = {https://pubs.rsc.org/en/content/articlelanding/2016/dt/c5dt03991f},
	doi = {10.1039/C5DT03991F},
	language = {en},
	number = {14},
	urldate = {2025-09-30},
	journal = {Dalton Transactions},
	publisher = {The Royal Society of Chemistry},
	author = {Morgan, Matthew M. and Piers, Warren E.},
	month = mar,
	year = {2016},
	pages = {5920--5924},
}

@article{holzmeierPhotoionizationPyrolysis14Azaborinine2014,
	title = {Photoionization and {Pyrolysis} of a 1,4-{Azaborinine}: {Retro}-{Hydroboration} in the {Cation} and {Identification} of {Novel} {Organoboron} {Ring} {Systems}},
	volume = {20},
	copyright = {© 2014 WILEY-VCH Verlag GmbH \& Co. KGaA, Weinheim},
	issn = {1521-3765},
	shorttitle = {Photoionization and {Pyrolysis} of a 1,4-{Azaborinine}},
	url = {https://onlinelibrary.wiley.com/doi/abs/10.1002/chem.201402884},
	doi = {10.1002/chem.201402884},
	language = {en},
	number = {31},
	urldate = {2025-09-30},
	journal = {Chemistry – A European Journal},
	author = {Holzmeier, Fabian and Lang, Melanie and Hemberger, Patrick and Bodi, Andras and Schäfer, Marius and Dewhurst, Rian D. and Braunschweig, Holger and Fischer, Ingo},
	year = {2014},
	note = {\_eprint: https://chemistry-europe.onlinelibrary.wiley.com/doi/pdf/10.1002/chem.201402884},
	pages = {9683--9692},
}

@article{dewarNewHeteroaromaticCompounds1968,
	title = {New {Heteroaromatic} {Compounds}. {XXVIII}. {Preparation} and {Properties} of 10,9-{Borazaronaphthalene}},
	volume = {90},
	issn = {0002-7863},
	url = {https://doi.org/10.1021/ja01010a601},
	doi = {10.1021/ja01010a601},
	number = {8},
	urldate = {2025-09-30},
	journal = {Journal of the American Chemical Society},
	publisher = {American Chemical Society},
	author = {Dewar, Michael and Jones, Richard},
	month = apr,
	year = {1968},
	pages = {2137--2144},
}

@article{dewar624NewHeteroaromatic1958,
	title = {624. {New} heteroaromatic compounds. {Part} {I}. 9-{Aza}-10-boraphenanthrene},
	issn = {0368-1769},
	url = {https://pubs.rsc.org/en/content/articlelanding/1958/jr/jr9580003073},
	doi = {10.1039/JR9580003073},
	language = {en},
	number = {0},
	urldate = {2025-09-30},
	journal = {Journal of the Chemical Society (Resumed)},
	publisher = {The Royal Society of Chemistry},
	author = {Dewar, M. J. S. and Kubba, Ved P. and Pettit, R.},
	month = jan,
	year = {1958},
	pages = {3073--3076},
}

@article{linLongRangeCorrectedHybrid2013,
	title = {Long-{Range} {Corrected} {Hybrid} {Density} {Functionals} with {Improved} {Dispersion} {Corrections}},
	volume = {9},
	issn = {1549-9618},
	url = {https://doi.org/10.1021/ct300715s},
	doi = {10.1021/ct300715s},
	number = {1},
	urldate = {2025-06-18},
	journal = {Journal of Chemical Theory and Computation},
	publisher = {American Chemical Society},
	author = {Lin, You-Sheng and Li, Guan-De and Mao, Shan-Ping and Chai, Jeng-Da},
	month = jan,
	year = {2013},
	pages = {263--272},
}

@article{nazariUltrafastDynamicsPolycyclic2019,
	title = {Ultrafast dynamics in polycyclic aromatic hydrocarbons: the key case of conical intersections at higher excited states and their role in the photophysics of phenanthrene monomer},
	volume = {21},
	issn = {1463-9084},
	shorttitle = {Ultrafast dynamics in polycyclic aromatic hydrocarbons},
	url = {https://pubs.rsc.org/en/content/articlelanding/2019/cp/c9cp03147b},
	doi = {10.1039/C9CP03147B},
	language = {en},
	number = {31},
	urldate = {2025-02-05},
	journal = {Physical Chemistry Chemical Physics},
	publisher = {The Royal Society of Chemistry},
	author = {Nazari, M. and Bösch, C. D. and Rondi, A. and Francés-Monerris, A. and Marazzi, M. and Lognon, E. and Gazzetto, M. and Langenegger, S. M. and Häner, R. and Feurer, T. and Monari, A. and Cannizzo, A.},
	month = aug,
	year = {2019},
	pages = {16981--16988},
}

@article{zhangElectronDelocalizationLowlying2023,
	title = {Between electron delocalization and low-lying excited states of {BN}-doped aromatic hydrocarbons},
	volume = {825},
	issn = {0009-2614},
	url = {https://www.sciencedirect.com/science/article/pii/S0009261423003202},
	doi = {10.1016/j.cplett.2023.140615},
	urldate = {2024-03-26},
	journal = {Chemical Physics Letters},
	author = {Zhang, Chen and Chrostowska, Anna and Liu, Shih-Yuan and Karamanis, Panaghiotis and Otero, Nicolás},
	month = aug,
	year = {2023},
	pages = {140615},
}

@misc{frischetal.Gaussian16RevisionA032016,
	title = {Gaussian-16 {Revision} {A}.03},
	author = {Frisch et al., M. J.},
	year = {2016},
	note = {Gaussian Inc. Wallingford CT},
}

@article{bieblSwitchingShapesReversible2026,
	title = {Switching {Shapes}: {Reversible} {Three} {Species} {Photoisomerization} of {Substituted} 1,2-{Dihydro}-1,2-azaborinines},
	copyright = {https://creativecommons.org/licenses/by/4.0/},
	issn = {0002-7863, 1520-5126},
	shorttitle = {Switching {Shapes}},
	url = {https://pubs.acs.org/doi/10.1021/jacs.5c20667},
	doi = {10.1021/jacs.5c20667},
	language = {en},
	urldate = {2026-02-13},
	journal = {Journal of the American Chemical Society},
	author = {Biebl, Sonja M. and Lienert, Jonas N. and Müller, Adrian J. and Ströbele, Markus and Dreuw, Andreas and Wachtveitl, Josef and Bettinger, Holger F.},
	month = feb,
	year = {2026},
	pages = {jacs.5c20667},
}

@article{bannwarthGFN2xTBAnAccurateBroadly2019,
	title = {{GFN2}-{xTB}—{An} {Accurate} and {Broadly} {Parametrized} {Self}-{Consistent} {Tight}-{Binding} {Quantum} {Chemical} {Method} with {Multipole} {Electrostatics} and {Density}-{Dependent} {Dispersion} {Contributions}},
	volume = {15},
	issn = {1549-9618},
	url = {https://doi.org/10.1021/acs.jctc.8b01176},
	doi = {10.1021/acs.jctc.8b01176},
	number = {3},
	urldate = {2025-09-04},
	journal = {Journal of Chemical Theory and Computation},
	publisher = {American Chemical Society},
	author = {Bannwarth, Christoph and Ehlert, Sebastian and Grimme, Stefan},
	month = mar,
	year = {2019},
	pages = {1652--1671},
}

@article{luComprehensiveElectronWavefunction2024,
	title = {A comprehensive electron wavefunction analysis toolbox for chemists, {Multiwfn}},
	volume = {161},
	issn = {0021-9606},
	url = {https://doi.org/10.1063/5.0216272},
	doi = {10.1063/5.0216272},
	number = {8},
	urldate = {2024-11-14},
	journal = {The Journal of Chemical Physics},
	author = {Lu, Tian},
	month = aug,
	year = {2024},
	pages = {082503},
}

@article{henkelmanClimbingImageNudged2000,
	title = {A climbing image nudged elastic band method for finding saddle points and minimum energy paths},
	volume = {113},
	issn = {0021-9606},
	url = {https://aip.scitation.org/doi/10.1063/1.1329672},
	doi = {10.1063/1.1329672},
	number = {22},
	urldate = {2018-10-27},
	journal = {J. Chem. Phys.},
	author = {Henkelman, Graeme and Uberuaga, Blas P. and Jónsson, Hannes},
	month = nov,
	year = {2000},
	pages = {9901--9904},
}

@article{kendallElectronAffinitiesFirstrow1992,
	title = {Electron affinities of the first‐row atoms revisited. {Systematic} basis sets and wave functions},
	volume = {96},
	issn = {0021-9606},
	url = {https://aip.scitation.org/doi/10.1063/1.462569},
	doi = {10.1063/1.462569},
	number = {9},
	urldate = {2022-11-21},
	journal = {The Journal of Chemical Physics},
	publisher = {American Institute of Physics},
	author = {Kendall, Rick A. and Dunning, Thom H. and Harrison, Robert J.},
	month = may,
	year = {1992},
	pages = {6796--6806},
}

@article{liuLeastStableIsomer2017,
	title = {The {Least} {Stable} {Isomer} of {BN} {Naphthalene}: {Toward} {Predictive} {Trends} for the {Optoelectronic} {Properties} of {BN} {Acenes}},
	volume = {139},
	issn = {0002-7863},
	shorttitle = {The {Least} {Stable} {Isomer} of {BN} {Naphthalene}},
	url = {https://doi.org/10.1021/jacs.7b02661},
	doi = {10.1021/jacs.7b02661},
	number = {17},
	urldate = {2022-09-19},
	journal = {Journal of the American Chemical Society},
	publisher = {American Chemical Society},
	author = {Liu, Zhiqiang and Ishibashi, Jacob S. A. and Darrigan, Clovis and Dargelos, Alain and Chrostowska, Anna and Li, Bo and Vasiliu, Monica and Dixon, David A. and Liu, Shih-Yuan},
	month = may,
	year = {2017},
	note = {Number: 17},
	pages = {6082--6085},
}

@article{negriVibronicStructureEmission1996,
	title = {Vibronic structure of the emission spectra from single vibronic levels of the {S1} manifold in naphthalene: {Theoretical} simulation},
	volume = {104},
	issn = {0021-9606},
	shorttitle = {Vibronic structure of the emission spectra from single vibronic levels of the {S1} manifold in naphthalene},
	url = {https://aip.scitation.org/doi/abs/10.1063/1.471054},
	doi = {10.1063/1.471054},
	number = {10},
	urldate = {2022-09-28},
	journal = {The Journal of Chemical Physics},
	publisher = {American Institute of Physics},
	author = {Negri, Fabrizia and Zgierski, Marek Z.},
	month = mar,
	year = {1996},
	note = {Number: 10},
	pages = {3486--3500},
}

@article{robeyPrioriCalculationsVibronic1977,
	title = {A priori calculations on vibronic coupling in the {1B2u}  {1Ag} (3200 Å) and higher transitions of naphthalene},
	volume = {23},
	issn = {0301-0104},
	url = {https://www.sciencedirect.com/science/article/pii/0301010477890034},
	doi = {10.1016/0301-0104(77)89003-4},
	language = {en},
	number = {2},
	urldate = {2022-10-31},
	journal = {Chemical Physics},
	author = {Robey, M. J. and Ross, I. G. and Southwood-Jones, Rosalind V. and Strickler, S. J.},
	month = jul,
	year = {1977},
	note = {Number: 2},
	pages = {207--216},
}

@article{zgierskiCommentsVibronicIntensities1977,
	title = {Some comments on vibronic intensities of naphthalene fluorescence spectra from single vibronic levels},
	volume = {47},
	issn = {0009-2614},
	url = {https://www.sciencedirect.com/science/article/pii/0009261477850252},
	doi = {10.1016/0009-2614(77)85025-2},
	language = {en},
	number = {3},
	urldate = {2022-10-31},
	journal = {Chemical Physics Letters},
	author = {Zgierski, Marek Z.},
	month = may,
	year = {1977},
	note = {Number: 3},
	pages = {499--502},
}

@article{henkelmanImprovedTangentEstimate2000,
	title = {Improved tangent estimate in the nudged elastic band method for finding minimum energy paths and saddle points},
	volume = {113},
	issn = {0021-9606, 1089-7690},
	url = {http://aip.scitation.org/doi/10.1063/1.1323224},
	doi = {10.1063/1.1323224},
	language = {en},
	number = {22},
	urldate = {2018-08-22},
	journal = {J. Chem. Phys.},
	author = {Henkelman, Graeme and Jónsson, Hannes},
	month = dec,
	year = {2000},
	pages = {9978--9985},
}

@article{singhEnergeticsOptimalMolecular2021,
	title = {Energetics and optimal molecular packing for singlet fission in {BN}-doped perylenes: electronic adiabatic state basis screening},
	volume = {23},
	shorttitle = {Energetics and optimal molecular packing for singlet fission in {BN}-doped perylenes},
	url = {https://pubs.rsc.org/en/content/articlelanding/2021/cp/d1cp01762d},
	doi = {10.1039/D1CP01762D},
	language = {en},
	number = {31},
	urldate = {2022-02-03},
	journal = {Physical Chemistry Chemical Physics},
	publisher = {Royal Society of Chemistry},
	author = {Singh, Anurag and Humeniuk, Alexander and S. Röhr, Merle I.},
	year = {2021},
	pages = {16525--16536},
}

@article{zhuGeodesicInterpolationReaction2019,
	title = {Geodesic interpolation for reaction pathways},
	volume = {150},
	issn = {0021-9606},
	url = {https://doi.org/10.1063/1.5090303},
	doi = {10.1063/1.5090303},
	number = {16},
	urldate = {2023-09-19},
	journal = {The Journal of Chemical Physics},
	author = {Zhu, Xiaolei and Thompson, Keiran C. and Martínez, Todd J.},
	month = apr,
	year = {2019},
	pages = {164103},
}

@article{shiozakiBAGELBrilliantlyAdvanced2018,
	title = {{BAGEL}: {Brilliantly} {Advanced} {General} {Electronic}-structure {Library}},
	volume = {8},
	copyright = {© 2017 Wiley Periodicals, Inc.},
	issn = {1759-0884},
	shorttitle = {{BAGEL}},
	url = {https://onlinelibrary.wiley.com/doi/abs/10.1002/wcms.1331},
	doi = {10.1002/wcms.1331},
	language = {en},
	number = {1},
	urldate = {2023-09-19},
	journal = {WIREs Computational Molecular Science},
	author = {Shiozaki, Toru},
	year = {2018},
	note = {\_eprint: https://onlinelibrary.wiley.com/doi/pdf/10.1002/wcms.1331},
	pages = {e1331},
}
\bibliographystyle{rsc} %the RSC's .bst file

\end{document}

\pagestyle{plain}
\thispagestyle{plain}
\fancypagestyle{plain}{
%%%HEADER%%%
\renewcommand{\headrulewidth}{0pt}
}
%%%END OF HEADER%%%

%%%PAGE SETUP - Please do not change any commands within this section%%%
\makeFNbottom
\makeatletter
\renewcommand\LARGE{\@setfontsize\LARGE{15pt}{17}}
\renewcommand\Large{\@setfontsize\Large{12pt}{14}}
\renewcommand\large{\@setfontsize\large{10pt}{12}}
\renewcommand\footnotesize{\@setfontsize\footnotesize{7pt}{10}}
\makeatother

\renewcommand{\thefootnote}{\fnsymbol{footnote}}
\renewcommand\footnoterule{\vspace*{1pt}% 
\color{cream}\hrule width 3.5in height 0.4pt \color{black}\vspace*{5pt}} 
\setcounter{secnumdepth}{5}

\makeatletter 
\renewcommand\@biblabel[1]{#1}            
\renewcommand\@makefntext[1]% 
{\noindent\makebox[0pt][r]{\@thefnmark\,}#1}
\makeatother 
\renewcommand{\figurename}{\small{Fig.}~}
\sectionfont{\sffamily\Large}
\subsectionfont{\normalsize}
\subsubsectionfont{\bf}
\setstretch{1.125} %In particular, please do not alter this line.
\setlength{\skip\footins}{0.8cm}
\setlength{\footnotesep}{0.25cm}
\setlength{\jot}{10pt}
\titlespacing*{\section}{5pt}{4pt}{20pt}
\titlespacing*{\subsection}{5pt}{15pt}{1pt}
\titleformat{\section}{\huge\bfseries}{\thesection}{1em}{}          
%%%END OF PAGE SETUP%%%

%%%FOOTER%%%
%\fancyfoot{}

% \fancyfoot[RO]{\footnotesize{\sffamily{1--\pageref{LastPage} ~\textbar  \hspace{2pt}\thepage}}}
% \fancyfoot[LE]{\footnotesize{\sffamily{\thepage~\textbar\hspace{4.65cm} 1--\pageref{LastPage}}}}
%\fancyhead{}
\renewcommand{\headrulewidth}{0pt} 
\renewcommand{\footrulewidth}{0pt}
\setlength{\arrayrulewidth}{1pt}
\setlength{\columnsep}{6.5mm}
\setlength\bibsep{1pt}
%%%END OF FOOTER%%%

%%%FIGURE SETUP - please do not change any commands within this section%%%
\makeatletter 
\newlength{\figrulesep} 
\setlength{\figrulesep}{0.5\textfloatsep} 

\newcommand{\topfigrule}{\vspace*{-1pt}% 
\noindent{\color{cream}\rule[-\figrulesep]{\columnwidth}{1.5pt}} }

\newcommand{\botfigrule}{\vspace*{-2pt}% 
\noindent{\color{cream}\rule[\figrulesep]{\columnwidth}{1.5pt}} }

\newcommand{\dblfigrule}{\vspace*{-1pt}% 
\noindent{\color{cream}\rule[-\figrulesep]{\textwidth}{1.5pt}} }

\makeatother
%%%END OF FIGURE SETUP%%%

%%%TITLE, AUTHORS AND ABSTRACT%%%

\begin{center}
      {\Huge\bfseries
    Supplementary Information
    }
\end{center}
\begin{center}
  % Title

  \vspace{1.0cm}
  {\LARGE\bfseries
    Asymmetric Planar-to-Dewar Isomerisation in BN-Doped Naphthalene: Mechanistic Implications for Molecular Solar Thermal Storage\textsuperscript{\dag}
  }

  \vspace{0.5cm}

  % Authors
  {Michael Bühler,\textit{$^{a}$} Merle I. S. Röhr$^{\ast}$\textit{$^{a,b}$} 
  }

  \vspace{0.3cm}

  % Affiliation
  {\small\itshape
    \textsuperscript{a}Institute of Physical and Theoretical Chemistry, \\ University of Würzburg, Emil-Fischer-Str. 42, 97074 Würzburg, Germany \\
    \textsuperscript{b}Center for Nanosystems Chemistry, \\
    Universität Würzburg, Theodor-Boveri-Weg, 97074 Würzburg, Germany.
  }

  \vspace{0.3cm}

  % Email
  {\small
    \texttt{merle.roehr@uni-wuerzburg.de}
  }

  \vspace{0.2cm}

\end{center}
\thispagestyle{empty}
  
%%%END OF TITLE, AUTHORS AND ABSTRACT%%%

%%%FONT SETUP - please do not change any commands within this section
\renewcommand*\rmdefault{bch}\normalfont\upshape
\rmfamily

{
\tableofcontents 
\addtocontents{toc}{\protect\thispagestyle{empty}}
\pagenumbering{arabic}
}
\clearpage

%%%END OF FOOTNOTES%%%

%%%MAIN TEXT%%%%

\section{NICS structure}

\begin{figure}[h]
    \centering
    \includegraphics[]{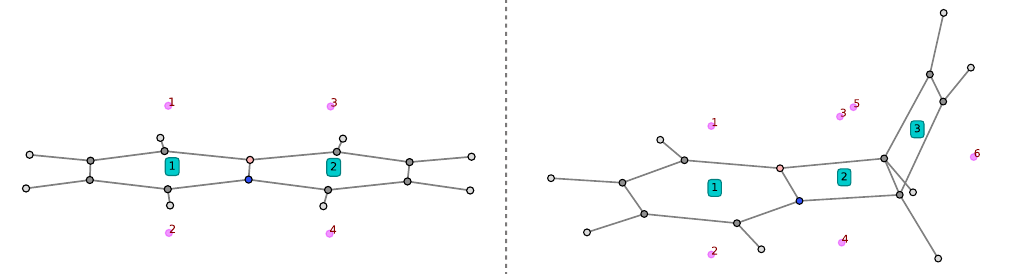}
    \caption{Structure representations derived from \texttt{pyAroma\,5}\cite{wangPyAromaIntuitiveGraphical2024} showing the probe points employed for the evaluation of $\mathrm{NICS}_{\perp}$. The planar structure is shown on the left, whereas the Dewar form is shown on the right. The rings are numbered consecutively to facilitate assignment of the individual aromatic subunits, and the positions at which $\mathrm{NICS}_{\perp}$ was evaluated are indicated for each ring. Nitrogen atoms are highlighted in blue and boron atoms in pink.}
    \label{fig:pyAroma}
\end{figure}

\section{Lewis structure of \bnBenz and derivatives}

\begin{figure}[h]
    \centering
    \includegraphics[]{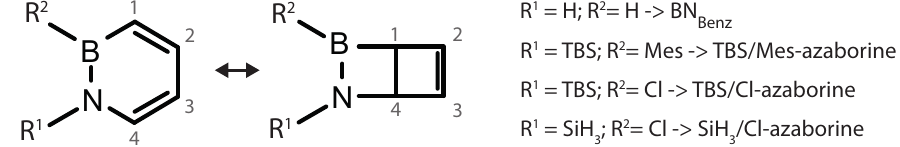}
    \caption{Lewis structures of the planar \bnBenz isomer (left) and the Dewar valence isomer (right). The atom numbering follows the convention used for \bnNaph (see Fig.~1). The trivial names used throughout the text are indicated for clarity.}
    \label{fig:Lewis_BN_Benz}
\end{figure}

\clearpage
\section{NEB pathway for \ce{SiH3}/\ce{Cl}-azaborine and \ce{SiH3}/\ce{Cl}-benzene}

\begin{figure}[h]
    \centering
    \includegraphics[]{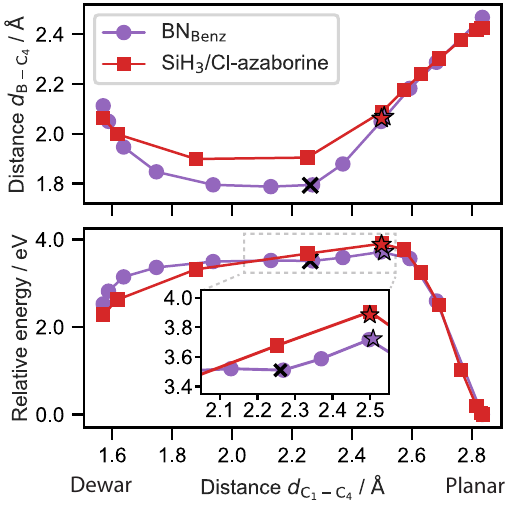}
    \caption{Nudged elastic band pathway connecting the planar and Dewar forms for two \bnBenz systems, the 1,2-dihydro parent compound and the disubstituted derivative bearing a \ce{SiH3} substituent at nitrogen and a \ce{Cl} substituent at boron. Both panels are plotted along the conversion pathway defined by the \(d_{\ce{C1-C4}}\) distance on the x-axis. The top panel shows the \(\ce{B-C4}\) distance, denoted as \(d_{\ce{B-C4}}\), along the transformation. The bottom panel shows the corresponding relative energy profile along the same conversion pathway, with an inset magnifying the region around the local minimum. The \bnBenz profile exhibits a shallow local minimum in this region, whereas this minimum disappears upon substitution with \ce{SiH3} and \ce{Cl}, resulting in a monotonic energy decrease and indicating destabilisation of the intermediate along the ground-state pathway. The positions of the corresponding optimised transition states are marked by star symbols, while the identified optimised local minimum is indicated by a black cross.}
    \label{fig:BN_Benz_Rest}
\end{figure}

\begin{figure}[h!]
    \centering
    \includegraphics[]{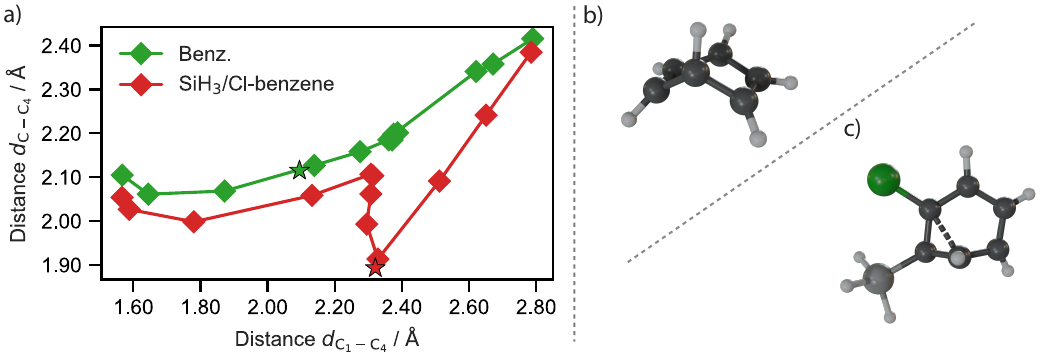}
    \caption{Figure X. Comparison of the nudged elastic band (NEB) pathways for benzene and \ce{SiH3}/\ce{Cl}-substituted benzene. (a) NEB pathway projected onto the two key \ce{C-C} bond distances, showing the progression of the transformation in both systems. The x-axis corresponds to the \ce{C1-C4} distance, while the y-axis denotes the distance between the carbon atom bearing the chlorine substituent and the  \ce{C4} atom. The notation follows Fig.~\ref{fig:Lewis_BN_Benz}. (b) Transition state geometry of benzene. (c) Transition state geometry of \ce{SiH3}/\ce{Cl}-substituted benzene.}
    \label{fig:Benz_Rest}
\end{figure}

\clearpage
\section{Structural comparison of the NEB transition state and the conical intersection of \Naph}

\begin{figure}[h]
    \centering
    \includegraphics[]{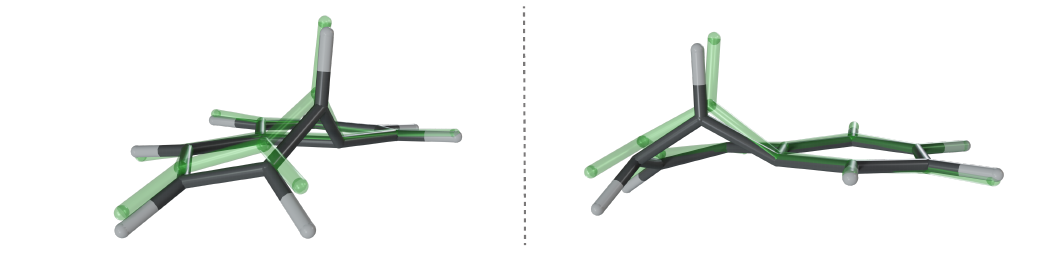}
    \caption{Structural comparison between the transition state for the planar to Dewar isomerisation of naphthalene and the minimum energy conical intersection along the nonradiative deactivation pathway, shown in green. The superimposed geometries allow a direct assessment of their similarity with respect to the out of plane distortion of the ring framework and the approach at the reactive boron carbon contact. The left panel shows the side view, while the right panel shows the front view.}
    \label{fig:placeholder}
\end{figure}

\clearpage
\section{Effects of substitution on \bnNaph}
\subsection{Natural transition orbitals for the S$_1$ of \bnNaph and the 1-dimethylamino-substituted derivative}

\begin{figure}[h!]
    \centering
    \includegraphics[]{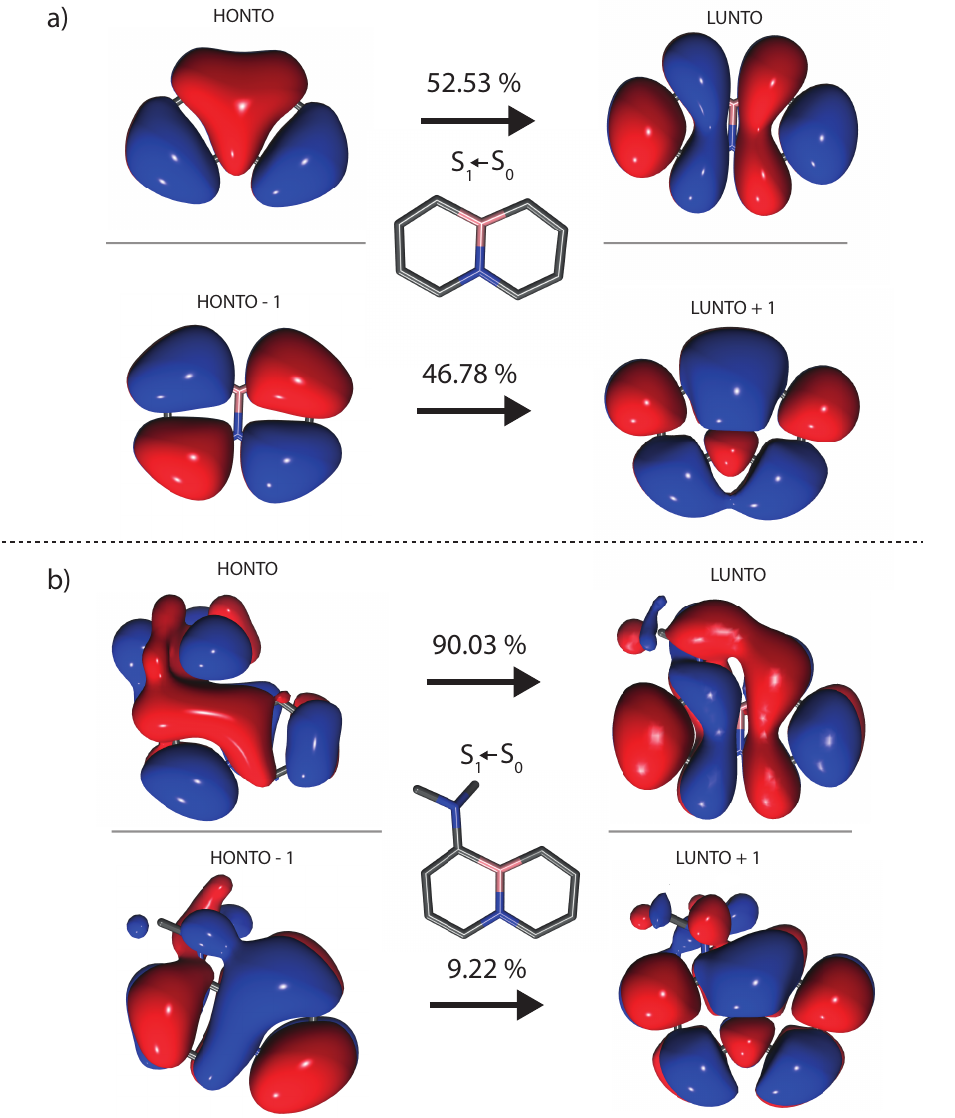}
    \caption{Natural transition orbitals (NTOs) associated with the S$_1$ excitation for (a) \bnNaph and (b) the 8-dimethylamino-substituted derivative, calculated at the $\omega$B97X-D3/aug-cc-pVDZ level of theory. For each compound, the leading hole and particle NTO pairs are shown together with their relative contributions to the excitation. Hydrogen atoms are omitted for clarity. All NTO isosurfaces were rendered using an isovalue of 0.02.}
    \label{fig:NTO}
\end{figure}

\clearpage

\subsection{Substituent effects on excitation energies and oscillator strength of \bnNaph derivatives}

\begin{table}[htbp]
\caption{Tabulated substituent- and position-dependent shifts of the S$_1$ excitation energy relative to \bnNaph\ for the planar and Dewar forms, given as $\Delta E_\mathrm{S1}^\mathrm{planar}$ and $\Delta E_\mathrm{S1}^\mathrm{Dewar}$, respectively. The corresponding oscillator strengths are listed as $f_\mathrm{osc}^\mathrm{planar}$ for the planar form and $f_\mathrm{osc}^\mathrm{Dewar}$ for the Dewar form.}
\centering
\scriptsize
\setlength{\tabcolsep}{3.5pt}
\renewcommand{\arraystretch}{1.15}
% Requires: \usepackage{booktabs,multirow,siunitx}
% Shifts are relative to HYDROGEN_REF: S1(R) = 4.667 eV, S1(P) = 4.863 eV.
% Reference oscillator strengths: fosc(R) = 0.003349, fosc(P) = 0.091380.

\begin{minipage}[t]{0.49\textwidth}

\centering
\begin{tabular}{ccSSSS}
\toprule
Rest & Position & {$\Delta E_\mathrm{S1}^\mathrm{planar}$ / eV} & {$f_\mathrm{osc}^\mathrm{planar}$} & {$\Delta E_\mathrm{S1}^\mathrm{Dewar}$ / eV} & {$f_\mathrm{osc}^\mathrm{Dewar}$} \\
\midrule
\multirow[c]{8}{*}{$-\mathrm{NO_2}$} & 1 & -0.769 & 0.0079 & -0.571 & 0.0004 \\
 & 2 & -0.556 & 0.0000 & -0.965 & 0.0001 \\
 & 3 & -0.813 & 0.0000 & -0.912 & 0.0000 \\
 & 4 & -0.917 & 0.0000 & -0.573 & 0.0005 \\
 & 5 & -0.916 & 0.0000 & -0.843 & 0.0378 \\
 & 6 & -0.813 & 0.0000 & -0.746 & 0.0001 \\
 & 7 & -0.556 & 0.0000 & -0.906 & 0.0001 \\
 & 8 & -0.671 & 0.0000 & -0.900 & 0.0001 \\
\midrule
\multirow[c]{8}{*}{$-\mathrm{CN}$} & 1 & -0.197 & 0.0644 & 0.035 & 0.0821 \\
 & 2 & -0.153 & 0.0075 & -0.058 & 0.0430 \\
 & 3 & -0.077 & 0.0358 & -0.065 & 0.0399 \\
 & 4 & -0.135 & 0.0481 & 0.021 & 0.0858 \\
 & 5 & -0.135 & 0.0481 & -0.238 & 0.1953 \\
 & 6 & -0.077 & 0.0358 & -0.004 & 0.0683 \\
 & 7 & -0.153 & 0.0075 & -0.289 & 0.0892 \\
 & 8 & -0.197 & 0.0645 & -0.271 & 0.2165 \\
\midrule
\multirow[c]{8}{*}{$-\mathrm{CF_3}$} & 1 & -0.017 & 0.0175 & 0.028 & 0.0822 \\
 & 2 & -0.016 & 0.0007 & 0.028 & 0.0805 \\
 & 3 & 0.039 & 0.0095 & -0.006 & 0.0798 \\
 & 4 & -0.009 & 0.0090 & 0.016 & 0.0888 \\
 & 5 & -0.009 & 0.0089 & -0.005 & 0.1178 \\
 & 6 & 0.039 & 0.0095 & 0.130 & 0.0705 \\
 & 7 & -0.016 & 0.0007 & -0.088 & 0.0898 \\
 & 8 & -0.017 & 0.0176 & -0.011 & 0.1330 \\
\midrule
\multirow[c]{8}{*}{$-\mathrm{Cl}$} & 1 & -0.094 & 0.0193 & -0.018 & 0.0765 \\
 & 2 & 0.002 & 0.0023 & 0.027 & 0.0812 \\
 & 3 & -0.092 & 0.0153 & -0.036 & 0.0747 \\
 & 4 & -0.077 & 0.0092 & -0.077 & 0.0862 \\
 & 5 & -0.077 & 0.0092 & -0.046 & 0.1222 \\
 & 6 & -0.092 & 0.0153 & -0.212 & 0.0743 \\
 & 7 & 0.002 & 0.0023 & 0.044 & 0.0898 \\
 & 8 & -0.093 & 0.0193 & -0.123 & 0.1279 \\
\midrule
\multirow[c]{8}{*}{$-\mathrm{F}$} & 1 & -0.006 & 0.0003 & -0.031 & 0.0795 \\
 & 2 & 0.056 & 0.0135 & 0.043 & 0.0983 \\
 & 3 & -0.074 & 0.0111 & -0.017 & 0.0895 \\
 & 4 & -0.025 & 0.0005 & -0.062 & 0.0837 \\
 & 5 & -0.025 & 0.0005 & 0.034 & 0.0817 \\
 & 6 & -0.074 & 0.0111 & -0.216 & 0.0963 \\
 & 7 & 0.056 & 0.0135 & 0.201 & 0.0865 \\
 & 8 & -0.006 & 0.0003 & -0.006 & 0.0777 \\
\bottomrule
\end{tabular}
\end{minipage}%
\hfill
\begin{minipage}[t]{0.49\textwidth}

\centering
\begin{tabular}{ccSSSS}
\toprule
Rest & Position & {$\Delta E_\mathrm{S1}^\mathrm{planar}$ / eV} & {$f_\mathrm{osc}^\mathrm{planar}$} & {$\Delta E_\mathrm{S1}^\mathrm{Dewar}$ / eV} & {$f_\mathrm{osc}^\mathrm{Dewar}$} \\
\midrule
\multirow[c]{8}{*}{$-\mathrm{CH_2CH_3}$} & 1 & -0.043 & 0.0073 & -0.035 & 0.0793 \\
 & 2 & 0.023 & 0.0048 & -0.006 & 0.0806 \\
 & 3 & -0.029 & 0.0042 & -0.023 & 0.0860 \\
 & 4 & -0.062 & 0.0070 & -0.012 & 0.0911 \\
 & 5 & -0.062 & 0.0071 & -0.040 & 0.1224 \\
 & 6 & -0.049 & 0.0067 & -0.141 & 0.0811 \\
 & 7 & 0.006 & 0.0024 & 0.032 & 0.0827 \\
 & 8 & -0.043 & 0.0073 & -0.023 & 0.1174 \\
\midrule
\multirow[c]{8}{*}{$-\mathrm{CH_3}$} & 1 & -0.037 & 0.0058 & -0.040 & 0.0821 \\
 & 2 & 0.013 & 0.0033 & 0.002 & 0.0876 \\
 & 3 & -0.050 & 0.0054 & -0.033 & 0.0836 \\
 & 4 & -0.064 & 0.0085 & -0.006 & 0.0913 \\
 & 5 & -0.064 & 0.0085 & -0.049 & 0.1145 \\
 & 6 & -0.050 & 0.0054 & -0.155 & 0.0830 \\
 & 7 & 0.013 & 0.0033 & 0.060 & 0.0818 \\
 & 8 & -0.037 & 0.0058 & -0.013 & 0.1022 \\
\midrule
\multirow[c]{8}{*}{$-\mathrm{OCH_3}$} & 1 & -0.272 & 0.0461 & -0.048 & 0.0801 \\
 & 2 & 0.006 & 0.0273 & 0.021 & 0.0942 \\
 & 3 & -0.095 & 0.0123 & -0.078 & 0.0595 \\
 & 4 & -0.100 & 0.0065 & -0.065 & 0.0849 \\
 & 5 & -0.082 & 0.0042 & -0.136 & 0.1180 \\
 & 6 & -0.105 & 0.0136 & -0.205 & 0.0863 \\
 & 7 & -0.036 & 0.0086 & 0.146 & 0.0709 \\
 & 8 & -0.145 & 0.0192 & -0.325 & 0.1191 \\
\midrule
\multirow[c]{8}{*}{$-\mathrm{N(CH_3)_2}$} & 1 & -0.463 & 0.1094 & -0.050 & 0.0786 \\
 & 2 & -0.011 & 0.0027 & -0.047 & 0.0665 \\
 & 3 & -0.037 & 0.0051 & -0.079 & 0.0308 \\
 & 4 & -0.170 & 0.0228 & -0.079 & 0.0876 \\
 & 5 & -0.171 & 0.0227 & -0.302 & 0.1884 \\
 & 6 & -0.041 & 0.0066 & -0.528 & 0.0633 \\
 & 7 & 0.026 & 0.0027 & 0.182 & 0.0353 \\
 & 8 & -0.459 & 0.1081 & -0.586 & 0.1472 \\
\bottomrule
\end{tabular}
\end{minipage}
\end{table}

\clearpage

\subsection{Substituent effects on bond length \ce{B-C4} of \bnNaph derivatives}

\begin{table}[htbp]
\caption{Minimum \ce{B-C} bond length between the boron atom and the carbon atom at position 4 along the conversion pathway from the planar to the Dewar form.}
\centering
\small
\setlength{\tabcolsep}{6pt}
\renewcommand{\arraystretch}{1.15}
% Requires: \usepackage{booktabs,multirow,siunitx}

\begin{minipage}[t]{0.48\textwidth}
\centering
\begin{tabular}{ccS}
\toprule
Rest & Position & {$d_\mathrm{B-C4,min}$ / \AA} \\
\midrule
\multirow[c]{8}{*}{$-\mathrm{NO_2}$} & 1 & 1.926 \\
 & 2 & 1.887 \\
 & 3 & 1.931 \\
 & 4 & 2.113 \\
 & 5 & 1.854 \\
 & 6 & 1.857 \\
 & 7 & 1.814 \\
 & 8 & 1.853 \\
\midrule
\multirow[c]{8}{*}{$-\mathrm{CN}$} & 1 & 1.875 \\
 & 2 & 1.859 \\
 & 3 & 1.852 \\
 & 4 & 2.101 \\
 & 5 & 1.842 \\
 & 6 & 1.849 \\
 & 7 & 1.831 \\
 & 8 & 1.836 \\
\midrule
\multirow[c]{8}{*}{$-\mathrm{Cl}$} & 1 & 1.845 \\
 & 2 & 1.889 \\
 & 3 & 1.834 \\
 & 4 & 1.867 \\
 & 5 & 1.854 \\
 & 6 & 1.838 \\
 & 7 & 1.853 \\
 & 8 & 1.829 \\
\midrule
\multirow[c]{8}{*}{$-\mathrm{F}$} & 1 & 1.826 \\
 & 2 & 1.915 \\
 & 3 & 2.141 \\
 & 4 & 1.900 \\
 & 5 & 1.857 \\
 & 6 & 1.827 \\
 & 7 & 1.854 \\
 & 8 & 1.822 \\
\midrule
\multirow[c]{1}{*}{$-\mathrm{H}$} & Ref & 1.837 \\
\bottomrule
\end{tabular}
\end{minipage}%
\hfill
\begin{minipage}[t]{0.48\textwidth}
\centering
\begin{tabular}{ccS}
\toprule
Rest & Position & {$d_\mathrm{B-C,min}$ / \AA} \\
\midrule
\multirow[c]{8}{*}{$-\mathrm{CH_2CH_3}$} & 1 & 1.805 \\
 & 2 & 1.821 \\
 & 3 & 1.819 \\
 & 4 & 1.902 \\
 & 5 & 1.854 \\
 & 6 & 1.837 \\
 & 7 & 1.856 \\
 & 8 & 1.855 \\
\midrule
\multirow[c]{8}{*}{$-\mathrm{CH_3}$} & 1 & 1.831 \\
 & 2 & 1.844 \\
 & 3 & 1.827 \\
 & 4 & 1.877 \\
 & 5 & 1.851 \\
 & 6 & 1.836 \\
 & 7 & 1.849 \\
 & 8 & 1.855 \\
\midrule
\multirow[c]{8}{*}{$-\mathrm{OCH_3}$} & 1 & 2.091 \\
 & 2 & 1.848 \\
 & 3 & 2.089 \\
 & 4 & 2.044 \\
 & 5 & 1.878 \\
 & 6 & 1.833 \\
 & 7 & 1.871 \\
 & 8 & 1.860 \\
\midrule
\multirow[c]{8}{*}{$-\mathrm{N(CH_3)_2}$} & 1 & 2.128 \\
 & 2 & 1.858 \\
 & 3 & 2.058 \\
 & 4 & 2.135 \\
 & 5 & 1.859 \\
 & 6 & 1.842 \\
 & 7 & 1.906 \\
 & 8 & 1.829 \\
\bottomrule
\end{tabular}
\end{minipage}

\end{table}

\clearpage

\subsection{Classification of Conversion Pathways and NEB-TS Profiles}

\begin{figure}[h]
    \centering
    \includegraphics[]{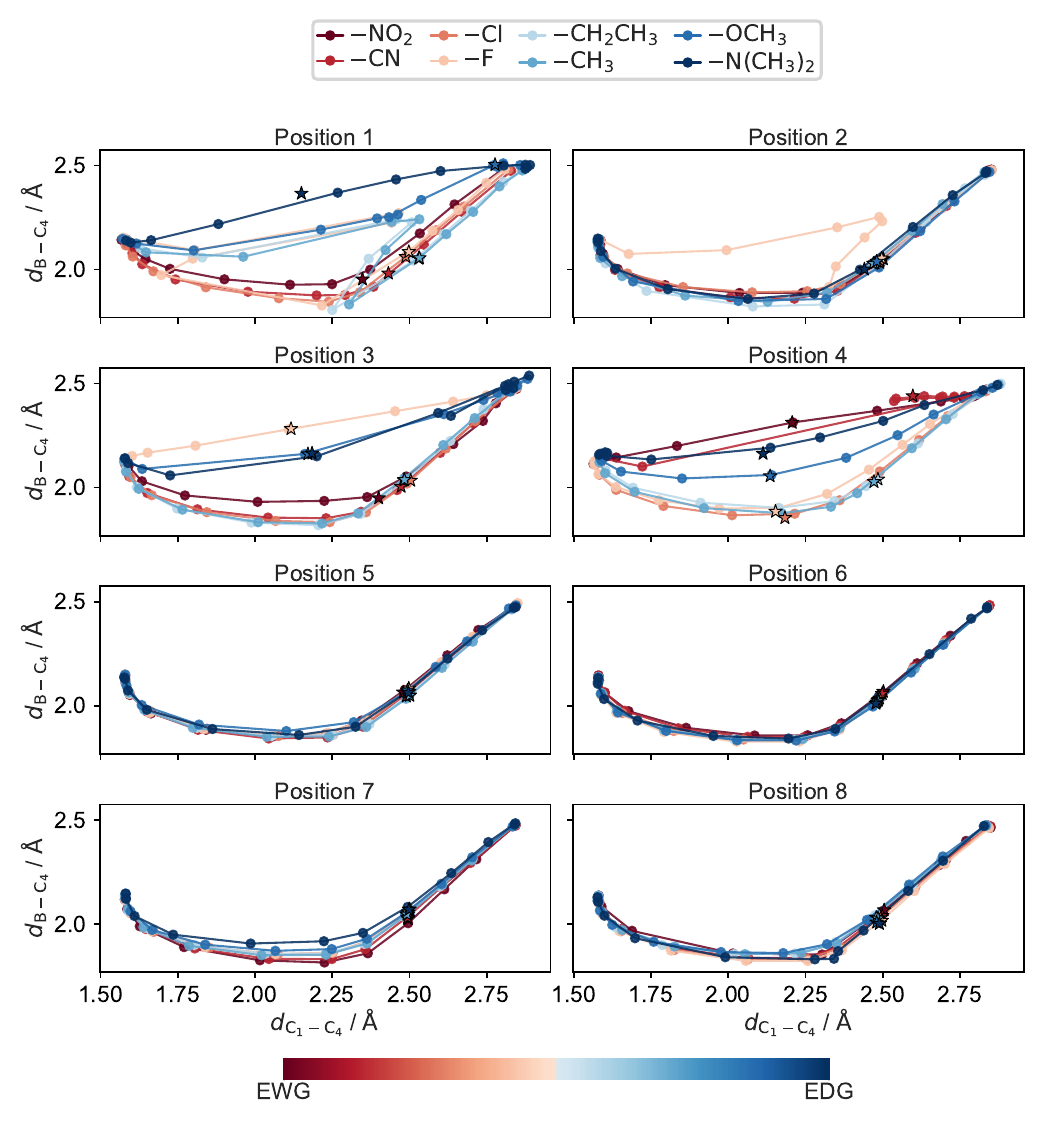}
    \caption{Calculated conversion pathways from the Dewar to the planar configuration of substituted \bnNaph\ derivatives with substituents at carbon positions 1--8. The conversion pathway is shown as the distance of the newly forming \ce{C1-C4} bond plotted against the \ce{B-C4} distance, illustrating the structural changes along the isomerization pathway. Different substituents and substitution positions clearly influence the course of the pathway. Optimized transition states are marked by colored stars corresponding to the respective substituents. All data were obtained using NEB-TS calculations at the $\omega$B97X-D3/aug-cc-pVDZ level of theory.}
    \label{fig:sub_position_all}
\end{figure}

\begin{table}[h]
\centering
\renewcommand{\arraystretch}{1.2}
\begin{tabular}{l p{2.7cm} p{3.7cm} p{2.7cm}}
\hline
\textbf{Substituent} & \textbf{Path 1: \bnNaph} & \textbf{Path 2: Naphthalene} & \textbf{Path 3: double} \\
\hline
Chloro (\ce{-Cl})          & 1--8                 & --     & --   \\
Cyano (\ce{-CN})           & 1--3, 5--7                       & --   & 4 \\
Nitro (\ce{-NO2})           & 1--3, 5--8              & --  & 4  \\
Fluoro (\ce{-F})          & 4--8                  & 1, 2 & 3  \\
Methyl (\ce{-CH3})          & 2--8             & 1   & --  \\
Ethyl (\ce{-CH2CH3})           & 2--8            & 1   & --   \\
Methoxy (\ce{-OCH3})         & 2, 6--8                  & 1  & 3, 4  \\
Dimethylamino (\ce{-N(CH3)2})   & 2, 5--8               & --   & 1, 3, 4 \\
\hline
\end{tabular}
\caption{Classification of the conversion pathways for \bnNaph\ derivatives functionalized at positions 1--8. Path~1 follows a \bnNaph-like pathway, whereas Path~2 shows a naphthalene-like pathway. Path~3 is characterized by a concerted upward displacement of both sides. Cases exhibiting features intermediate between Path~1 and Path~2 are assigned to a mixed category.}
\end{table}

\clearpage

\subsection{Activation and storage energies}

\begin{figure}[h]
    \centering
    \includegraphics[]{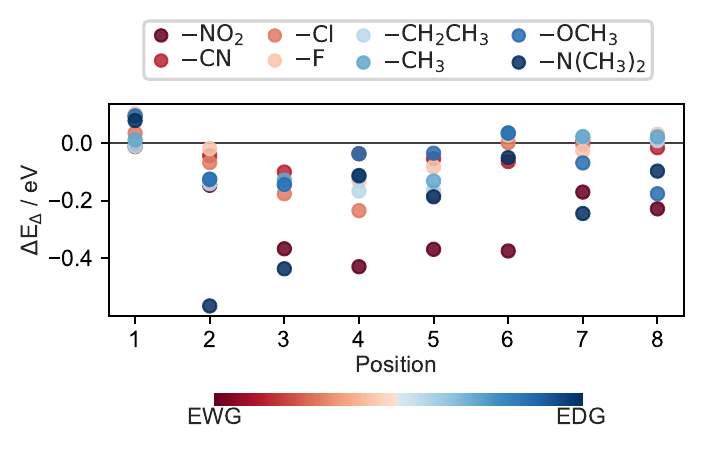}
    \caption{Relative energies $\Delta E_{\Delta}$ of the planar and Dewar forms of substituted \bnNaph derivatives as a function of substitution position, given with respect to \bnNaph. For \bnNaph, the reference value obtained at the DFT level is $\Delta E = 2.43$~eV between the planar and Dewar forms.  The corresponding absolute values are provided in Tab. \ref{tab:storage}.}
    \label{fig:relativ_energy}
\end{figure}

\begin{figure}[h!]
    \centering
    \includegraphics[]{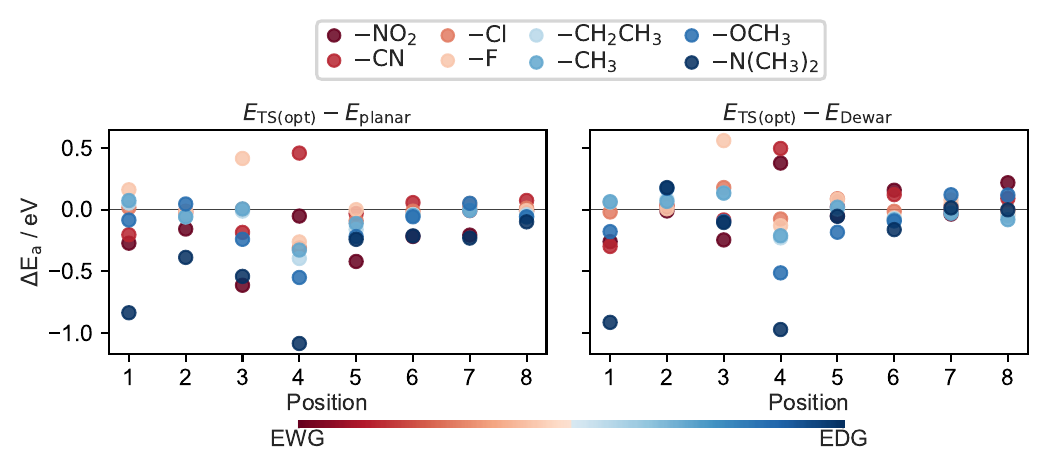}
    \caption{Relative activation energies for the conversion between the planar and Dewar forms of substituted \bnNaph\ derivatives as a function of substitution position, given with respect to \bnNaph. Left: barrier from the planar structure to the transition state. Right: barrier from the Dewar structure to the transition state. For \bnNaph, the corresponding DFT reference barriers are 4.24~eV and 1.80~eV from the planar and Dewar structures to the transition state, respectively. The corresponding absolute values are provided in Tab.~\ref{tab:ts}.} 
    \label{fig:act_energy}
\end{figure}

\begin{table}[h!]
\caption{Tabulated values of the energy differences between the planar and Dewar forms for substitution at positions 1--8, corresponding to the data shown in Fig. \ref{fig:relativ_energy}.}
\label{tab:storage}
\centering
\small
\setlength{\tabcolsep}{6pt}
\renewcommand{\arraystretch}{1.15}
% Requires: \usepackage{booktabs,multirow,siunitx}

\begin{minipage}[t]{0.48\textwidth}
\centering
\begin{tabular}{ccS}
\toprule
 & Position & {$E_\mathrm{Dewar}-E_\mathrm{planar}$ / eV} \\
\midrule
\multirow[c]{8}{*}{$-\mathrm{NO_2}$} & 1 & 2.424 \\
 & 2 & 2.290 \\
 & 3 & 2.069 \\
 & 4 & 2.007 \\
 & 5 & 2.068 \\
 & 6 & 2.062 \\
 & 7 & 2.266 \\
 & 8 & 2.208 \\
\midrule
\multirow[c]{8}{*}{$-\mathrm{CN}$} & 1 & 2.532 \\
 & 2 & 2.393 \\
 & 3 & 2.336 \\
 & 4 & 2.399 \\
 & 5 & 2.382 \\
 & 6 & 2.372 \\
 & 7 & 2.442 \\
 & 8 & 2.420 \\
\midrule
\multirow[c]{8}{*}{$-\mathrm{Cl}$} & 1 & 2.471 \\
 & 2 & 2.369 \\
 & 3 & 2.260 \\
 & 4 & 2.202 \\
 & 5 & 2.258 \\
 & 6 & 2.438 \\
 & 7 & 2.432 \\
 & 8 & 2.453 \\
\midrule
\multirow[c]{8}{*}{$-\mathrm{F}$} & 1 & 2.538 \\
 & 2 & 2.417 \\
 & 3 & 2.292 \\
 & 4 & 2.305 \\
 & 5 & 2.354 \\
 & 6 & 2.464 \\
 & 7 & 2.409 \\
 & 8 & 2.467 \\
\bottomrule
\end{tabular}
\end{minipage}%
\hfill
\begin{minipage}[t]{0.48\textwidth}
\centering
\begin{tabular}{ccS}
\toprule
Rest & Position & {$E_\mathrm{Dewar}-E_\mathrm{planar}$ / eV} \\
\midrule
\multirow[c]{8}{*}{$-\mathrm{CH_2CH_3}$} & 1 & 2.425 \\
 & 2 & 2.295 \\
 & 3 & 2.292 \\
 & 4 & 2.270 \\
 & 5 & 2.271 \\
 & 6 & 2.469 \\
 & 7 & 2.455 \\
 & 8 & 2.445 \\
\midrule
\multirow[c]{8}{*}{$-\mathrm{CH_3}$} & 1 & 2.447 \\
 & 2 & 2.312 \\
 & 3 & 2.308 \\
 & 4 & 2.321 \\
 & 5 & 2.305 \\
 & 6 & 2.472 \\
 & 7 & 2.459 \\
 & 8 & 2.458 \\
\midrule
\multirow[c]{8}{*}{$-\mathrm{OCH_3}$} & 1 & 2.531 \\
 & 2 & 2.310 \\
 & 3 & 2.292 \\
 & 4 & 2.399 \\
 & 5 & 2.401 \\
 & 6 & 2.471 \\
 & 7 & 2.367 \\
 & 8 & 2.260 \\
\midrule
\multirow[c]{8}{*}{$-\mathrm{N(CH_3)_2}$} & 1 & 2.515 \\
 & 2 & 1.871 \\
 & 3 & 2.000 \\
 & 4 & 2.324 \\
 & 5 & 2.250 \\
 & 6 & 2.386 \\
 & 7 & 2.192 \\
 & 8 & 2.339 \\
\bottomrule
\end{tabular}
\end{minipage}

\end{table}

\begin{table}[h!]
\caption{Tabulated activation energies for the conversion between the planar and Dewar forms of substituted \bnNaph\ derivatives as a function of substitution position, corresponding to the barriers shown in Fig. \ref{fig:act_energy}.}
\centering
\small
\setlength{\tabcolsep}{6pt}
\label{tab:ts}
\renewcommand{\arraystretch}{1.15}
% Requires: \usepackage{booktabs,multirow,siunitx}
\begin{minipage}[t]{0.48\textwidth}
\centering
\begin{tabular}{ccSS}
\toprule
Rest & Position & {$E_\mathrm{TS}-E_\mathrm{planar}$ / eV} & {$E_\mathrm{TS}-E_\mathrm{Dewar}$ / eV} \\
\midrule
\multirow[c]{8}{*}{$-\mathrm{NO_2}$} & 1 & 3.969 & 1.545 \\
 & 2 & 4.083 & 1.794 \\
 & 3 & 3.628 & 1.559 \\
 & 4 & 4.189 & 2.181 \\
 & 5 & 3.821 & 1.753 \\
 & 6 & 4.023 & 1.961 \\
 & 7 & 4.034 & 1.768 \\
 & 8 & 4.230 & 2.022 \\
\midrule
\multirow[c]{8}{*}{$-\mathrm{CN}$} & 1 & 4.037 & 1.506 \\
 & 2 & 4.225 & 1.833 \\
 & 3 & 4.056 & 1.720 \\
 & 4 & 4.699 & 2.300 \\
 & 5 & 4.209 & 1.826 \\
 & 6 & 4.298 & 1.926 \\
 & 7 & 4.233 & 1.791 \\
 & 8 & 4.315 & 1.895 \\
\midrule
\multirow[c]{8}{*}{$-\mathrm{Cl}$} & 1 & 4.257 & 1.785 \\
 & 2 & 4.210 & 1.841 \\
 & 3 & 4.243 & 1.983 \\
 & 4 & 3.931 & 1.729 \\
 & 5 & 4.150 & 1.893 \\
 & 6 & 4.229 & 1.790 \\
 & 7 & 4.268 & 1.837 \\
 & 8 & 4.253 & 1.800 \\
\midrule
\multirow[c]{8}{*}{$-\mathrm{F}$} & 1 & 4.402 & 1.863 \\
 & 2 & 4.228 & 1.811 \\
 & 3 & 4.655 & 2.364 \\
 & 4 & 3.979 & 1.674 \\
 & 5 & 4.241 & 1.887 \\
 & 6 & 4.202 & 1.738 \\
 & 7 & 4.291 & 1.882 \\
 & 8 & 4.233 & 1.766 \\
\bottomrule
\end{tabular}
\end{minipage}%
\hfill
\begin{minipage}[t]{0.48\textwidth}
\centering
\begin{tabular}{ccSS}
\toprule
Rest & Position & {$E_\mathrm{TS}-E_\mathrm{planar}$ / eV} & {$E_\mathrm{TS}-E_\mathrm{Dewar}$ / eV} \\
\midrule
\multirow[c]{8}{*}{$-\mathrm{CH_2CH_3}$} & 1 & 4.294 & 1.869 \\
 & 2 & 4.182 & 1.887 \\
 & 3 & 4.229 & 1.937 \\
 & 4 & 3.845 & 1.575 \\
 & 5 & 4.079 & 1.808 \\
 & 6 & 4.200 & 1.731 \\
 & 7 & 4.238 & 1.782 \\
 & 8 & 4.199 & 1.754 \\
\midrule
\multirow[c]{8}{*}{$-\mathrm{CH_3}$} & 1 & 4.315 & 1.867 \\
 & 2 & 4.183 & 1.871 \\
 & 3 & 4.247 & 1.939 \\
 & 4 & 3.913 & 1.593 \\
 & 5 & 4.129 & 1.824 \\
 & 6 & 4.194 & 1.721 \\
 & 7 & 4.234 & 1.775 \\
 & 8 & 4.181 & 1.723 \\
\midrule
\multirow[c]{8}{*}{$-\mathrm{OCH_3}$} & 1 & 4.157 & 1.626 \\
 & 2 & 4.287 & 1.976 \\
 & 3 & 4.000 & 1.708 \\
 & 4 & 3.691 & 1.292 \\
 & 5 & 4.022 & 1.621 \\
 & 6 & 4.183 & 1.712 \\
 & 7 & 4.292 & 1.925 \\
 & 8 & 4.183 & 1.922 \\
\midrule
\multirow[c]{8}{*}{$-\mathrm{N(CH_3)_2}$} & 1 & 3.405 & 0.891 \\
 & 2 & 3.854 & 1.983 \\
 & 3 & 3.700 & 1.700 \\
 & 4 & 3.156 & 0.832 \\
 & 5 & 4.000 & 1.750 \\
 & 6 & 4.028 & 1.642 \\
 & 7 & 4.010 & 1.818 \\
 & 8 & 4.142 & 1.803 \\
\bottomrule
\end{tabular}
\end{minipage}

\end{table}

\clearpage

%%%END OF MAIN TEXT%%%


\balance


%%%REFERENCES%%%
\bibliography{references} %You need to replace "rsc" on this line with the name of your .bib file
\bibliographystyle{rsc} %the RSC's .bst file